\input ./style/arxiv-general.cfg
\documentclass[aos,MSNbibl,nameyear,dvips]{arximspdf}
\makeatletter
   \@ifpackageloaded{graphicx}{}{\usepackage{graphicx}}
\makeatother
\usepackage{dcolumn}

\doi{10.1214/14-AOS1307}
\volume{43}
\issue{4}
\pubyear{2015}
\firstpage{1498}
\lastpage{1534}
\docsubty{FLA}

\makeatletter
\newcolumntype{d}[1]{D{.}{.}{#1}}

\def\sfrac#1#2{#1/#2}

\def\vafrac#1#2{(#1)/(#2)}

\newcommand{\lleft}{\left}
\newcommand{\rrvert}{\vert}
\newcommand{\rright}{\right}
\newcommand{\llvert}{\vert}
\renewcommand{\mid}{|}
\newcommand{\begcomment}{\iffalse}

\newtheorem{teo}{Theorem}[section]
\newtheorem{lem}{Lemma}[section]
\newtheorem{coro}{Corollary}[section]
\newtheorem{prop}{Proposition}[section]
\newproclaim{rem}{Remark}
\def\argmin{\mathop{\operatorname{argmin}}}

\def\argmin{\mathop{\operatorname{argmin}}}
\def\tr{\operatorname{tr}}
\def\vec{\operatorname{vec}}
\def\sgn{\operatorname{sign}}
\def\sym{\operatorname{sym}}
\def\var{\operatorname{var}}
\def\Rq{\operatorname{Rq}}
\def\Err{\operatorname{Err}}

\newcommand{\Xb}{\mathbf{X}}
\newcommand{\Zb}{\mathbf{Z}}
\newcommand{\Ub}{\mathbf{U}}
\newcommand{\Yb}{\mathbf{Y}}
\newcommand{\Wb}{\mathbf{W}}
\newcommand{\bB}{\mathbf{B}}
\newcommand{\bx}{\mathbf{x}}
\newcommand{\bv}{\mathbf{v}}
\newcommand{\bY}{\mathbf{Y}}
\newcommand{\bS}{\mathbf{S}}
\newcommand{\bZ}{\mathbf{Z}}
\newcommand{\bQ}{\mathbf{Q}}
\newcommand{\bD}{\mathbf{D}}
\newcommand{\bC}{\mathbf{C}}
\newcommand{\bI}{\mathbf{I}}
\newcommand{\bK}{\mathbf{K}}
\newcommand{\bw}{\mathbf{w}}
\newcommand{\bA}{\mathbf{A}}
\newcommand{\bzero}{{\mathbf0}}

\newcommand{\bdelta}{\bolds{\delta}}
\newcommand{\bbeta}{\bolds{\beta}}
\newcommand{\bSigma}{\bolds{\Sigma}}
\newcommand{\hbSigma}{\widehat{\bSigma}}
\newcommand{\bmu}{\bolds{\mu}}
\newcommand{\hbmu}{\widehat{\bmu}}
\newcommand{\bOmega}{\bolds{\Omega}}
\newcommand{\hbOmega}{\widehat{\bOmega}}
\newcommand{\hbdelta}{\widehat{\bdelta}}
\newcommand{\hbQ}{\widehat{\bQ}}
\newcommand{\hbx}{\widehat{\bx}}
\makeatother

\begin{document}
\begin{frontmatter}

\title{QUADRO: A supervised dimension reduction method via Rayleigh
quotient optimization}
\runtitle{Dimension reduction via Rayleigh optimization}

\begin{aug}
\author[A]{\fnms{Jianqing}~\snm{Fan}\corref{}\thanksref{M1,T1}\ead[label=e1]{jqfan@princeton.edu}},
\author[B]{\fnms{Zheng Tracy}~\snm{Ke}\thanksref{M2,T1}\ead[label=e2]{zke@galton.uchicago.edu}},
\author[A]{\fnms{Han}~\snm{Liu}\thanksref{M1,T2}\ead[label=e3]{hanliu@princeton.edu}}
\and
\author[A]{\fnms{Lucy}~\snm{Xia}\thanksref{M1,T1}\ead[label=e4]{lxia@princeton.edu}}
\runauthor{Fan, Ke, Liu and Xia}
\affiliation{Princeton University\thanksmark{M1} and University of Chicago\thanksmark{M2}}
\address[A]{J. Fan\\
H. Liu\\
L. Xia\\
Department of Operations Research\\
\quad and Financial Engineering\\
Princeton University\\
Princeton, New Jersey 08544\\
USA\\
\printead{e1}\\
\phantom{E-mail: }\printead*{e3}\\
\phantom{E-mail: }\printead*{e4}}
\address[B]{Z. Ke\\
Department of Statistics\\
University of Chicago\\
Chicago, Illinois 60637\\
USA\\
\printead{e2}}
\end{aug}
\thankstext{T1}{Supported in part by NSF Grants DMS-12-06464 and
DMS-14-06266 and NIH Grants \mbox{R01-GM100474} and R01-GM072611.}
\thankstext{T2}{Supported in part by NSF Grants III-1116730, NSF
III-1332109, an NIH sub-award and
a FDA sub-award from Johns Hopkins University and an NIH-subaward from
Harvard University.}

%
\received{\smonth{11} \syear{2013}}
%
\revised{\smonth{12} \syear{2014}}

%
\begin{abstract}
We propose a novel Rayleigh quotient based sparse quadratic dimension
reduction method---named QUADRO (\underline{Qua}dratic \underline
{D}imension \underline{R}eduction via Rayleigh \underline
{O}ptimization)---for analyzing high-\break dimensional data. Unlike in the
linear setting where Rayleigh quotient optimization coincides with
classification, these two problems are very different under nonlinear
settings. In this paper, we clarify this difference and show that
Rayleigh quotient optimization may be of independent scientific
interests. One major challenge of Rayleigh quotient optimization is
that the variance of quadratic statistics involves all fourth
cross-moments of predictors, which are infeasible to compute for
high-dimensional applications and may accumulate too many stochastic
errors. This issue is resolved by considering a family of elliptical
models. Moreover, for heavy-tail distributions, robust estimates of
mean vectors and covariance matrices are employed to guarantee uniform
convergence in estimating nonpolynomially many parameters, even though
only the fourth moments are assumed.
Methodologically, QUADRO is based on elliptical models which allow us
to formulate the Rayleigh quotient maximization as a convex
optimization problem. Computationally, we propose an efficient
linearized augmented Lagrangian method to solve the constrained
optimization problem. Theoretically, we provide explicit rates of
convergence in terms of Rayleigh quotient under both Gaussian and
general elliptical models.
Thorough numerical results on both synthetic and real datasets are also
provided to back up our theoretical results.
\end{abstract}

%
\begin{keyword}[class=AMS]
\kwd[Primary ]{62H30}
\kwd[; secondary ]{62G20}
\end{keyword}
\begin{keyword}
\kwd{Classification}
\kwd{dimension reduction}
\kwd{quadratic discriminant analysis}
\kwd{Rayleigh quotient}
\kwd{oracle inequality}
\end{keyword}
\end{frontmatter}

\section{Introduction} \label{secintro}

Rapid developments of imaging technology, microarray data studies and
many other applications call for the analysis of high-dimensional
binary-labeled data. We consider the problem of finding a ``nice''
projection $f\dvtx \mathbb{R}^d\to\mathbb{R}$ that embeds all data into
the real line. A projection such as $f$ has applications in many
statistical problems for analyzing high-dimensional binary-labeled
data, including:
\begin{itemize}
\item \textit{Dimension reduction}: $f$ provides a data reduction tool for
people to visualize the high-dimensional data in a one-dimensional space.
\item \textit{Classification}: $f$ can be used to construct classification
rules. With a carefully chosen set $A\subset\mathbb{R}$, we can
classify a new data point $\bx\in\mathbb{R}^d$ by checking whether
or not $f(\bx)\in A$.
\item \textit{Feature selection}: when $f(\bx)$ only depends on a small
number of coordinates of~$\bx$, this projection selects just a few
features from numerous observed ones.
\end{itemize}

A natural question is what kind of $f$ is a ``nice'' projection? It
depends on the goal of statistical analysis. For classification, a good
$f$ should yield to a small classification error.
In feature selection, different criteria select distinct features, and
they may suit different real problems.
In this paper, we propose using the following criterion for finding $f$:

\begin{quote}
Under the mapping $f$, the data are as ``separable'' as
possible between two classes, and as ``coherent'' as possible within
each class.
\end{quote}

\noindent It can be formulated as to maximize the \emph{Rayleigh quotient} of
$f$. Suppose all data are drawn independently from a joint distribution
of $(\Xb, Y)$, where $\Xb\in\mathbb{R}^d$, and $Y\in\{0,1\}$ is
the label. The \emph{Rayleigh quotient} of $f$ is defined as
%
%
\begin{equation}
\label{Rq} \Rq(f) \equiv\frac{\var\{ \mathbb{E}[f(\Xb)\mid Y]\}}{\var
\{ f(\Xb)-\mathbb{E}[f(\Xb)\mid Y]\} }.
\end{equation}
Here, the numerator is the variance of $\Xb$ explained by the class
label, and the denominator is the remaining variance of $\Xb$.
Simple calculation shows that $\Rq(f)=\pi(1-\pi)R(f)$, where $\pi
\equiv\mathbb{P}(Y=0)$ and
%
%
\begin{equation}
\label{Rf} R(f) \equiv\frac{\{\mathbb{E}[f(\Xb)\mid Y=0] - \mathbb
{E}[f(\Xb
)\mid Y=1] \}^2}{\pi\var[f(\Xb)\mid Y=0] + (1-\pi)\var[f(\Xb)\mid Y=1]}.
\end{equation}
Our\vspace*{1pt} goal is to develop a data-driven procedure to find $\hat{f}$ such
that $\Rq(\hat{f})$ is large, and $\hat{f}$ is sparse in the sense
that it depends on few coordinates of $\Xb$.

The Rayleigh quotient, as a criterion for finding a projection $f$,
serves different purposes. First, for dimension reduction, it takes
care of both variance explanation and label explanation. In contrast,
methods such as principal component analysis (PCA) only consider
variance explanation. Second, when the data are normally distributed, a
monotone transform of the Rayleigh quotient approximates the
classification error; see Section~\ref{secclassification}. Therefore,
an $f$ with a large Rayleigh quotient enables us to construct nice
classification rules. In addition, it is a convex optimization to
maximize the Rayleigh quotient among linear and quadratic $f$ (see
Section~\ref{secmethod}), while minimizing the classification error
is not.
Third, with appropriate regularization, this criterion
provides a new feature selection tool for data analysis.

The criterion (\ref{Rq}), initially introduced by \citet{fisher1936use}
for classification, is known as Fisher's linear discriminant analysis
(LDA). In the literature of sufficient dimension reduction, the sliced
inverse regression (SIR) proposed by \citet{li1991sliced} can also be
formulated as maximizing (\ref{Rq}), where $Y$ can be any variable not
necessarily binary.
In both LDA and SIR, $f$ is restricted to be a linear function, and the
dimension $d$ cannot be larger than $n$. In this sense, our work
compares directly to various versions of LDA and SIR generalized to
nonlinear, high-dimensional settings.
We provide a more detailed comparison to the literature in Section~\ref
{secconclude}, but preview here the uniqueness of our work. First, we
consider a setting where $\Xb\mid Y$ has an elliptical distribution and
$f$ is a quadratic function, which allows us to derive a simplified
version of (\ref{Rq}) and gain extra statistical efficiency; see
Section~\ref{secformul} for details. This simplified version of
(\ref{Rq}) was never considered before. Furthermore, the assumption of
conditional elliptical distribution does not satisfy the requirement of
SIR and many other dimension reduction methods [\citet
{li1991sliced,cook1991comment}]. In Section~\ref{subsecobjective},
we explain the
motivation of the current setting. Second, we utilize robust estimators
of mean and covariance matrix, while many generalizations of LDA and
SIR are based on sample mean and sample covariance matrix. As shown in
Section~\ref{secestimation}, the robust estimators adapt better to
heavy tails on the data. It is worth noting that QUADRO only considers
the projection to a one-dimensional subspace. In contrast, more
sophisticated dimension reduction methods (e.g., the kernel SIR) are
able to find multiple projections $f_1,\ldots, f_m$ for $m>1$. This
reflects a tradeoff between modeling tractability and flexibility. More
specifically, QUADRO achieves better computational and theoretical
properties at the cost of sacrificing some flexibility.


\subsection{Rayleigh quotient and classification error} \label{subsecRqvsErr}

Many popular statistical methods for analyzing high-dimensional
binary-labeled data are based on classification error minimization,
which is closely related to the Rayleigh quotient maximization.
We summarize their connections and differences as follows:
\begin{longlist}[(a)]
\item[(a)] In an ``ideal'' setting where two classes follow
multivariate normal distributions with a common covariance matrix and
the class of linear functions $f$ is considered, the two criteria are
exactly the same, with one being a monotone transform of the other.
\item[(b)] In a ``relaxed'' setting where two classes follow
multivariate normal distributions but
with nonequal covariance matrices and the class of quadratic functions~$f$
(including linear functions as special cases) is considered, the
two criteria are closely related in the sense that a monotone transform
of the Rayleigh quotient is an approximation of the classification error.
\item[(c)] In other settings, the two criteria can be very different.
\end{longlist}
%
We now show (a) and (c), and will discuss (b) in Section~\ref
{secclassification}.

For each $f$, we define a family of classifiers $h_c(\bx)=I\{f(\bx)<
c\}$ indexed by $c$, where $I(\cdot)$ is the indicator function.~%
For each given $c$, we define the classification error of $h_{c}$ to be
$\operatorname{err}(h_c)\equiv\mathbb{P}(h_c(\Xb)\neq Y)$. The
classification error of $f$ is then defined by
\[
\Err(f) \equiv\min_{c\in\mathbb{R}} \bigl\{ \operatorname{err}(h_c) \bigr\}.
\]
Most existing classification procedures aim at finding a data-driven
projection $\hat{f}$ such that $\Err(\hat{f})$ is small (the
threshold $c$ is usually easy to choose). Examples include linear
discriminant analysis (LDA) and its variations in high dimensions
[e.g., \citet{SCRDA,FAIR,LPD,Shao11,witten2011penalized,ROAD,Coda}],
quadratic discriminant analysis (QDA), support vector machine (SVM),
logistic regression, boosting, etc.

We now compare $\Rq(f)$ and $\Err(f)$. Let $\pi= \mathbb{P}(Y=0)$,
$\bmu_1=\mathbb{E}(\Xb\mid Y=0)$, $\bSigma_1=\operatorname{cov}(\Xb
\mid Y=0)$, $\bmu_2=\mathbb{E}(\Xb\mid Y=1)$ and $\bSigma
_2=\operatorname
{cov}(\Xb\mid Y=1)$.
We consider linear functions $\{f(\bx)=\mathbf{a}^{\top}\bx+b\dvtx  \mathbf
{a}\in\mathbb{R}^d, b\in\mathbb{R}\}$,
and write $\Rq(\mathbf{a})=\Rq(\mathbf{a}^{\top}\bx)$, $\Err(\mathbf
{a})=\Err(\mathbf{a}^{\top}\bx)$ for short.
By direct calculation, when the two classes have a common covariance matrix
$\bSigma$,
\[
\Rq(\mathbf{a}) = \pi(1-\pi) \frac{[\mathbf{a}^{\top}(\bmu_1-\bmu
_2)]^2}{\mathbf{a}^{\top}\bSigma\mathbf{a}}.
\]
Hence, the optimal $\mathbf{a}_R=\bSigma^{-1}(\bmu_1-\bmu_2)$. On the
other hand, when data follow multivariate normal distributions, the
optimal classifier is
$h^*(\bx)=I\{\mathbf{a}_E^{\top}\bx< c\}$, where $\mathbf{a}_E=\bSigma
^{-1}(\bmu_1-\bmu_2)$ and $c=\tfrac{1}{2}\bmu_1^{\top}\bSigma
^{-1}\bmu_1 - \tfrac{1}{2}\bmu_2^{\top}\bSigma^{-1}\bmu_2 + \log
(\tfrac{1-\pi}{\pi})$.
It is observed that $\mathbf{a}_R=\mathbf{a}_E$ and the two criteria
are the
same. In fact, for all vectors $\mathbf{a}$ such that $\mathbf{a}^{\top
}(\bmu_1-\bmu_2)>0$,
\[
\Err(\mathbf{a}) = 1 - \Phi\biggl(\frac{1}{2} \biggl[ \frac{\Rq(\mathbf
{a})}{\pi(1-\pi)}
\biggr]^{1/2} \biggr),
\]
where $\Phi$ is the distribution function of a standard normal random
variable, and we fix $c=\mathbf{a}^{\top}(\bmu_1+\bmu_2)/2$. Therefore,
the classification error is a monotone transform of the Rayleigh quotient.

\begin{figure}

\includegraphics{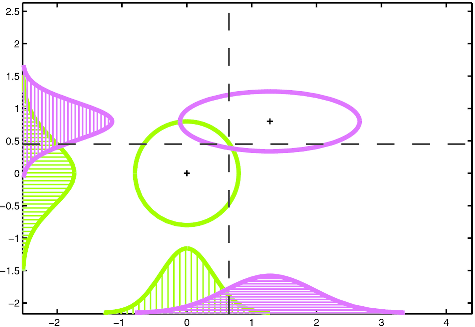}

\caption{An example in $\mathbb{R}^2$. The green and purple represent
class 1 and class 2, respectively. The ellipses are contours of
distributions. Probability densities after being projected to $X_1$ and
$X_2$ are also displayed. The dotted lines correspond to optimal
thresholds for classification using each feature.}
\label{figlinRQbayes}
\end{figure}

When we move away from these ideal assumptions, the above two criteria
can be very different. We illustrate this point using a bivariate
distribution, that is, $d=2$, with different covariance matrices.
Specifically, $\pi=0.55$, $\bmu_1=(0,0)^{\top}$, $\bmu_2=(1.28,
0.8)^{\top}$, $\bSigma_1=\operatorname{diag}(1,1)$ and $\bSigma
_2=\operatorname{diag}(3, 1/3)$. We still consider linear functions
$f(\bx)=\mathbf{a}^{\top}\bx$ but select only one out of the two
features, $X_1$ or $X_2$. Then the maximum Rayleigh quotients, by using
each of the two features alone, are 0.853 and 0.923, respectively,
whereas the minimum classification errors are 0.284 and 0.295,
respectively. As a result, under the criterion of maximizing Rayleigh
quotient, Feature~2 is selected, whereas under the criterion of
minimizing classification error, Feature~1 is selected. Figure~\ref
{figlinRQbayes} displays the distributions of data after being
projected to each of the two features. It shows that since data from
the second class has a much larger variability at Feature~1 than at
Feature~2, the Rayleigh quotient maximization favors Feature~2,
although Feature~1 yields a smaller classification error.

\subsection{Objective of the paper} \label{subsecobjective}

In this paper, we consider the Rayleigh quotient maximization problem
in the following setting:
\begin{itemize}
\item We consider sparse quadratic functions, that is, $f(\bx)=\bx
^{\top}\bOmega\bx- 2\bdelta^{\top}\bx$, where $\bOmega$ is a
sparse $d\times d$ symmetric matrix, and $\bdelta$ is a sparse
$d$-dimensional vector.
\item The two classes can have different covariance matrices.
\item Data from these two classes follow \textit{elliptical distributions}.
\item The dimension is large (it is possible that $d \gg n$).
\end{itemize}
%
Compared to Fisher's LDA, our setting has several new ingredients.
First, we go beyond linear classifiers to enhance flexibility. It is
well known that the linear classifiers are inefficient. For example,
when two classes have the same mean, linear classifiers perform no
better than random guesses. Instead of exploring arbitrary nonlinear
functions, we consider the class of quadratic functions so that the
Rayleigh quotient still has a nice parametric formulation, and at the
same time it helps identify interaction effects between features.
Second, we drop the requirement that the two classes share a common
covariance matrix, which is a critical condition for Fisher's rule and
many other high-dimensional classification methods [e.g., \citet
{FAIR,ROAD,LPD}].
In fact, by using quadratic discriminant functions, we take advantage
of the difference of covariance matrices between the two classes to
enhance classification power.
Third, we generalize multivariate normal distributions to
the elliptical family, which includes many heavy-tailed distributions,
such as multivariate $t$-distributions, Laplace distributions, and Cauchy\vspace*{1pt}
distributions. This family of distributions allows us to avoid
estimating all $O(d^4)$ fourth cross-moments of $d$ predictors in
computing the variance of quadratic statistics and hence overcomes the
computation and noise accumulation issues.

In our setting, Fisher's rule, that is, $\mathbf{a}_R=\bSigma^{-1}(\bmu
_1-\bmu_2)$, no longer maximizes the Rayleigh quotient. We propose a
new method, called quadratic dimension reduction via Rayleigh
optimization (QUADRO). It is a \emph{Rayleigh-quotient-oriented
procedure} and is a statistical tool for simultaneous dimension
reduction and feature selection. QUADRO has several properties. First,
it is a statistically efficient generalization of Fisher's linear
discriminant analysis to the quadratic setting.
A~naive generalization involves estimation of all fourth cross-moments
of the two underlying distributions. In contrast, QUADRO only requires
estimating a one-dimensional kurtosis parameter.
Second, QUADRO adopts rank-based estimators and robust $M$-estimators of
the covariance matrices and the means. Therefore, it is robust to
possibly heavy-tail distributions. Third, QUADRO can be formulated as a
convex programming and is computationally efficient.

Theoretically, we prove that under elliptical models, the Rayleigh
quotient of the estimated quadratic function $\hat{f}$ converges to
population maximum Rayleigh quotient at rate $O_p (s\sqrt{\log
(d)/n} )$, where $s$ is the number of important features (counting
both single terms and interaction terms).
In addition, we establish a connection between our method and quadratic
discriminant analysis (QDA) under elliptical models.


The rest of this paper is organized as follows. Section~\ref
{secformul} formulates Rayleigh quotient maximization as a convex
optimization problem. Section~\ref{secmethod} describes QUADRO.
Section~\ref{secestimation} discusses rank-based estimators and
robust $M$-estimators used in QUADRO. Section~\ref{secanalysis}
presents theoretical analysis. Section~\ref{secclassification}
discusses the application of QUADRO in elliptically distributed
classification problems. Section~\ref{secnumerical} contains
numerical studies. Section~\ref{secconclude}
concludes the paper.
All proofs are collected in Section~\ref{secproof}.

\subsubsection*{Notation}
For $0\leq q\leq\infty$, $\llvert \bv\rrvert _q$
denotes the
$L_q$-norm of a vector $\bv$, $\llvert \bA\rrvert _q$ denotes the elementwise
$L_q$-norm of a matrix $\bA$ and $\Vert\bA\Vert_q$ denotes the
matrix \mbox{$L_q$-}norm of $\bA$. When $q=2$, we omit the subscript $q$.
$\lambda_{\min}(\bA)$ and $\lambda_{\max}(\bA)$ denote the
minimum and maximum eigenvalues of $\bA$. $\det(\bA)$ denotes the
determinant of~$\bA$. Let $I(\cdot)$ be the indicator function:  for
any event $B$, $I(B)=1$ if $B$ happens and $I(B)=0$ otherwise. Let
$\sgn(\cdot)$ be the sign function, where $\sgn(u)=1$ when $u\geq0$
and $\sgn(u)=-1$ when $u<0$.

\section{Rayleigh quotient for quadratic functions} \label{secformul}

We first study the population form of Rayleigh quotient for an
arbitrary quadratic function. We show that it has a simplified form
under the elliptical family.

For a quadratic function
\[
Q(\Xb) = \Xb^{\top}\bOmega\Xb- 2\bdelta^{\top}\Xb,
\]
using (\ref{Rf}), its Rayleigh quotient is
%
%
\begin{equation}
\label{ROmega,delta} R(\bOmega,\bdelta) = \frac{ \{ \mathbb{E}[Q(\Xb
)\mid Y=0]-\mathbb
{E}[Q(\Xb)\mid Y=1] \}^2}{\pi\operatorname{var}[Q(\Xb)\mid Y=0] +
(1-\pi)\operatorname{var}[Q(\Xb)\mid Y=1]}
\end{equation}
up to a constant multiplier.
The Rayleigh quotient maximization can be expressed as
\[
\max_{(\bOmega, \bdelta)\dvtx  \bOmega=\bOmega^{\top}} R(\bOmega, \bdelta).
\]

\subsection{General setting}


Suppose $\mathbb{E}(\Zb)=\bmu$ and $\operatorname{cov}(\Zb
)=\bSigma$. By direct calculation,
\begin{eqnarray*}
\mathbb{E} \bigl[Q(\Zb) \bigr]&=& \tr(\bOmega\bSigma) + \bmu^{\top}
\bOmega\bmu- 2\bdelta^{\top}\bmu,
\\
\var\bigl[Q(\Zb) \bigr] &=& \mathbb{E} \bigl[\tr\bigl(\bOmega\Zb
\Zb^{\top} \bOmega\Zb\Zb^{\top} \bigr) \bigr] - 4\mathbb{E}
\bigl[\bdelta^{\top} \Zb\Zb^{\top}\bOmega\Zb\bigr]
\\
&&{} + 4
\bdelta^{\top}\bSigma\bdelta+4 \bigl(\bdelta^{\top}\bmu
\bigr)^2 - \bigl\{\mathbb{E} \bigl[Q(\Zb) \bigr] \bigr
\}^2.
\end{eqnarray*}
So $\mathbb{E}[Q(\Zb)]$ is a linear combination of the elements in $\{
\Omega(i,j), 1\leq i\leq j\leq d$; $\delta(i), 1\leq i\leq d\}$, and
$\var[Q(\Zb)]$ is a quadratic form of these elements. The
coefficients in $\mathbb{E}[Q(\Zb)]$ are functions of $\bmu$ and
$\bSigma$ only. However, the coefficients in $\var[Q(\Zb)]$ also
depend on all the fourth cross-moments of $\Zb$, and there are
$O(d^4)$ of them.

Let us define $M_1(\bOmega,\bdelta)=\mathbb{E}[Q(\Xb)\mid Y=0]$,
$L_1(\bOmega,\bdelta) = \operatorname{var}[Q(\Xb)\mid Y=0]$ and
$M_2(\bOmega,\bdelta)$, $L_2(\bOmega,\bdelta)$ similarly. Also, let
$\kappa= (1-\pi)/\pi$. We have
\[
R(\bOmega,\bdelta) = \frac{ [M_1(\bOmega,\bdelta)-M_2(\bOmega,\bdelta
)]^2}{L_1(\bOmega,\bdelta)+\kappa L_2(\bOmega,\bdelta)}.
\]
Therefore, both the numerator and denominator are quadratic
combinations of the elements in $\bOmega$ and $\bdelta$. We can stack
the $d(d+1)/2$ elements in $\bOmega$ (assuming it is symmetric) and
the $d$ elements in $\bdelta$ into a long vector $\bv$. Then
$R(\bOmega,\bdelta)$ can be written as
\[
R(\bv) = \frac{(\mathbf{a}^{\top}\bv)^2}{\bv^{\top}\bA\bv},
\]
where $\mathbf{a}$ is a $d'\times1$ vector, $\bA$ is a $d'\times d'$
positive semi-definite matrix and $d'=d(d+1)/2+d$. $\bA$ and $\mathbf{a}$
are determined by the coefficients in the denominator and numerator of
$R(\bOmega, \bdelta)$, respectively. Now, $\max_{(\bOmega,\bdelta
)}R(\bOmega,\bdelta)$ is equivalent to $\max_{\bv}R(\bv)$. It has
explicit solutions. For example, when $\bA$ is positive definite, the
function $R(\bv)$ is maximized at $\bv^* = \bA^{-1}\mathbf{a}$.
We can then reshape $\bv^*$ to get the desired $(\bOmega^*, \bdelta^*)$.


Practical implementation of the above idea is infeasible in high
dimensions as it involves $O(d^4)$ cross moments of $\bZ$. This not
only poses computational challenges, but also accumulates noise in the
estimation. Furthermore, good estimates of fourth moments usually
require the existence of eighth moments, which is not realistic for
many heavy tailed distributions. These problems can be avoided under
the elliptical family, as we now illustrate in the next subsection.


\subsection{Elliptical distributions}

The elliptical family contains multivariate distributions whose
densities have elliptical contours. It generalizes multivariate normal
distributions and inherits many of their nice properties.\vadjust{\goodbreak}

Given a $d\times1$ vector $\bmu$ and a $d\times d$ positive definite
matrix $\bSigma$, 
a random vector $\Zb$ that follows an elliptical distribution admits
%
%
\begin{equation}
\label{Zdecomposition} \Zb= \bmu+\xi\bSigma^{1/2} \Ub,
\end{equation}
where $\Ub$ is a random vector which follows the uniform distribution
on unit sphere $\mathcal{S}^{d-1}$, and $\xi$ is a nonnegative random
variable independent of $\Ub$.
Denote the elliptical distribution by $\mathcal{E}(\bmu,\bSigma,g)$,
where $g$ is the density of $\xi$. In this paper, we always assume
that $\mathbb{E}\xi^{4}<\infty$ and require that $\mathbb{E}(\xi
^2)=d$ for the model identifiability. Then $\bSigma$ is the covariance
matrix of $\Zb$.


%
\begin{prop} \label{propellipmeanvar}
Suppose $\Zb$ follows an elliptical distribution as in (\ref
{Zdecomposition}). Then 
\begin{eqnarray*}
\mathbb{E} \bigl[Q(\Zb) \bigr] &=& \tr(\bOmega\bSigma)+\bmu^{\top}\bOmega
\bmu-2\bmu^{\top}\bdelta,
\\
\var\bigl[Q(\Zb) \bigr] &=& 2(1+\gamma)\tr(
\bOmega\bSigma\bOmega\bSigma)+ \gamma\bigl[{\tr}(\bOmega\bSigma)
\bigr]^2 +4(\bOmega\bmu-\bdelta)^{\top
}\bSigma(\bOmega\bmu-
\bdelta),
\end{eqnarray*}
where $\gamma=\frac{E(\xi^4)}{d(d+2)}-1$ is the kurtosis parameter.
\end{prop}

The proof is given in the online supplementary material [\citet{QUADROsupp}]. The variance of $Q(\Zb)$ does not involve any fourth
cross-moments,\vspace*{1pt} but only the kurtosis parameter $\gamma$. For
multivariate normal distributions, $\xi^2$ follows a $\chi
^2$-distribution with $d$ degrees of freedom, and $\gamma=0$.
For multivariate $t$-distribution with degrees of freedom $\nu>4$, we
have $\gamma=2/(\nu-4)$.


\subsection{Rayleigh optimization}

We assume that the two classes both follow elliptical distributions:
$\Xb\mid(Y=0)\sim\mathcal{E}(\bmu_1,\bSigma_1, g_1)$ and $\Xb
\mid(Y=1)\sim\mathcal{E}(\bmu_2,\break \bSigma_2, g_2)$. To facilitate the
presentation, we assume the quantity $\gamma$ is the same for both
classes of conditional distributions. 
Let
%
%
\begin{eqnarray}
\label{ML1L2} M(\bOmega, \bdelta) &=& -\bmu^{\top}_1\bOmega
\bmu_1+\bmu^{\top
}_2\bOmega\bmu_2
+2( \bmu_1-\bmu_2)^{\top}\bdelta-\tr\bigl(
\bOmega( \bSigma_1-\bSigma_2) \bigr),\nonumber
\\
L_k(\bOmega, \bdelta) &=& 2(1+\gamma)\tr(\bOmega
\bSigma_k\bOmega\bSigma_k)+\gamma\bigl[{\tr}(\bOmega
\bSigma_k) \bigr]^2
\\
&&{} +4(\bOmega\bmu_k-
\bdelta)^{\top}\bSigma_k(\bOmega\bmu_k-
\bdelta),\nonumber
\end{eqnarray}
for $k=1$ and $2$. Combining (\ref{ROmega,delta}) with Proposition
\ref{propellipmeanvar}, we have
%
%
\begin{equation}
\label{defR} R(\bOmega,\bdelta) = \frac{[M(\bOmega,\bdelta
)]^2}{L_1(\bOmega,\bdelta)+ \kappa L_2(\bOmega,\bdelta)},
\end{equation}
where $\kappa= (1-\pi)/\pi$.

Note that if we multiply both $\bOmega$ and $\bdelta$ by a common
constant, $R(\bOmega,\bdelta)$ remains unchanged. Therefore,
maximizing $R(\bOmega, \bdelta)$ is equivalent to solving the
following constrained minimization problem:
%
%
\begin{equation}
\label{progunconstrain} \min_{(\bOmega,\bdelta)\dvtx  M(\bOmega,\bdelta
)=1,\bOmega=\bOmega
^{\top}} \bigl\{ L_1(\bOmega,
\bdelta)+\kappa L_2(\bOmega,\bdelta) \bigr\}.
\end{equation}
We call problem (\ref{progunconstrain}) the \textit{Rayleigh
optimization}. It is a convex problem whenever $\bSigma_1$ and
$\bSigma_2$ are both positive semi-definite.

The formulation of the Rayleigh optimization only involves the means
and covariance matrices, and the kurtosis parameter $\gamma$.
Therefore, if we know $\gamma$ (e.g., when we know which subfamily the
distributions belong to) and have good estimates $(\hbmu_1,\hbmu
_2,\hbSigma_1,\hbSigma_2)$, we can solve the empirical version of
(\ref{progunconstrain}) to obtain $(\widehat{\bOmega}, \widehat
{\bdelta})$, which is the main idea of QUADRO. In addition, (\ref
{progunconstrain}) is a convex problem, with a quadratic objective and
equality constraints. Hence it can be solved efficiently by many
optimization algorithms.

\section{Quadratic dimension reduction via Rayleigh optimization}
\label{secmethod}

Now, we formally introduce the QUADRO procedure. We fix a model
parameter $\gamma\geq0$. Let $\widehat{M}$, $\widehat{L}_1$ and
$\widehat{L}_2$ be the sample versions of $M, L_{1}, L_{2}$ in (\ref
{ML1L2}) by replacing $(\bmu_1,\bmu_2, \bSigma_1,\bSigma_2)$ with
their estimates. Details of these estimates will be given in
Section~\ref{secestimation}. Let $\widehat{\pi}=n_1/(n_1+n_2)$ and
$\kappa= \widehat{\pi}/(1-\widehat{\pi})$.
Given tuning parameters $\lambda_1>0$ and $\lambda_2>0$, we solve
%
%
\begin{equation}
\label{progquadro} \min_{(\bOmega,\bdelta)\dvtx  \widehat{M}(\bOmega,\bdelta
)=1, \bOmega
=\bOmega^{\top}} \bigl\{ \widehat{L}_1(
\bOmega,\bdelta)+\kappa\widehat{L}_2(\bOmega,\bdelta) +
\lambda_1 \llvert\bOmega\rrvert_1 + \lambda
_2 \llvert\bdelta\rrvert_1 \bigr\}.
\end{equation}



We propose a linearized augmented Lagrangian method
to solve (\ref{progquadro}).
To simplify the notation, we write $\widehat{L}=\widehat{L}_1+\kappa
\widehat{L}_2$, and omit the hat symbol on $M$ and $L$ when there is
no confusion. The optimization problem is then
\[
\min_{(\bOmega, \bdelta)\dvtx  M(\bOmega, \bdelta)=1, \bOmega=\bOmega
^{\top}} \bigl\{ L(\bOmega, \bdelta) + \lambda_1
\llvert\bOmega\rrvert_1 + \lambda_2 \llvert\bdelta
\rrvert_1 \bigr\}.
\]
For an algorithm parameter $\rho>0$, and a dual variable $\nu$, we
define the \textit{augmented Lagrangian} as
\[
F_{\rho}(\bOmega,\bdelta, \nu) = L(\bOmega,\bdelta) + \nu\bigl[M(
\bOmega,\bdelta)-1 \bigr] + (\rho/2) \bigl[M(\bOmega,\bdelta)-1
\bigr]^2.
\]
%
Using zero as the initial value, we iteratively update:
\begin{itemize}
\item$\bdelta^{(k)} = \argmin_{\bdelta} \{ F_{\rho}(\bOmega
^{(k-1)}, \bdelta, \nu^{(k-1)}) + \lambda_2 \llvert \bdelta\rrvert _1
\}$,\vspace*{1pt}
\item$\bOmega^{(k)} = \argmin_{\bOmega\dvtx  \bOmega=\bOmega^{\top
}}\{ F_{\rho}(\bOmega, \bdelta^{(k)}, \nu^{(k-1)}) + \lambda
_1 \llvert \bOmega\rrvert _1 \}$,\vspace*{1pt}
\item$\nu^{(k)} = \nu^{(k-1)} + \rho[M(\bOmega^{(k)},\bdelta^{(k)})-1]$.
\end{itemize}
Here, the first two steps are \textit{primal updates}, and the third step
is a \textit{dual update}.

First, we consider the update of $\bdelta$. When $\bOmega$ and $\nu$
are fixed, we can write
\[
F_\rho(\bOmega,\bdelta,\nu) = \bdelta^{\top}\bA\bdelta- 2
\bdelta^{\top}\mathbf{b}+ c_{\rho}(\bOmega,\nu),
\]
where
%
%
\begin{eqnarray}\label{alg-A,b}
\bA&=& 4(\bSigma_1+\kappa\bSigma_2) + 2\rho(
\bmu_1-\bmu_2) (\bmu_1-
\bmu_2)^{\top}, \nonumber
\\
\mathbf{b}&=& 4(\bSigma_1\bOmega\bmu_1+\kappa
\bSigma_2\bOmega\bmu_2)
\\
&&{} + \bigl[ \rho\tr\bigl(
\bOmega(\bSigma_1-\bSigma_2) \bigr) + \rho\bmu
_1^{\top}\bOmega\bmu_1 - \rho
\bmu_2^{\top}\bOmega\bmu_2 + (\rho-\nu) \bigr](
\bmu_1-\bmu_2),\hspace*{-30pt}
\nonumber
\end{eqnarray}
and $c_\rho(\bOmega, \nu)$ does not depend on $\bdelta$. Note that
$\bA$ is a positive semi-definite matrix. The update of $\bdelta$ is
indeed a Lasso problem.

Next, we consider the update of $\bOmega$. When $\bdelta$ and $\nu$
are fixed, $F_{\rho}(\bOmega,\bdelta,\nu)$ is a convex function of
$\bOmega$. We propose an approximate update step: we first
``linearize'' $F_\rho$ at $\bOmega=\bOmega^{(k-1)}$ to construct an
upper envelope $\bar{F}_\rho$, and then minimize this upper envelope.
In detail, at any $\bOmega=\bOmega_0$, we consider the following
upper bound of $F_\rho(\bOmega, \bdelta, \nu)$:
\begin{eqnarray*}
\bar{F}_\rho(\bOmega,\bdelta,\nu) &\equiv& F_\rho(
\bOmega_0, \bdelta,\nu) + \sum_{1\leq i\leq j\leq d}
\bigl[\Omega(i,j)-\Omega_0(i,j) \bigr]\frac{\partial F_\rho(\bOmega
_0,\bdelta,\nu)}{\partial
\Omega(i,j)}
\\
&&{} +\frac{\tau}{2}\sum_{1\leq i\leq j\leq d} \bigl[\Omega(i,j)-
\Omega_0(i,j) \bigr]^2,
\end{eqnarray*}
where $\tau$ is a large enough constant [e.g., we can take $\tau=\sum
_{1\leq i\leq j\leq d}\tfrac{\partial^2 F_{\rho}(\bOmega_0,\bdelta
,\nu)}{\partial\Omega(i,j)^2}$]. We then minimize $\bar{F}_\rho
(\bOmega,\bdelta,\nu)+\lambda_1\llvert \bOmega\rrvert _1$ to update
$\bOmega$.
This modified update step has an explicit solution,
\[
\Omega^*(i,j)= \mathcal{S} \biggl( \Omega_0(i,j) -
\frac{1}{\tau
}\frac{\partial F_\rho(\bOmega_0,\bdelta,\nu)}{\partial\Omega
(i,j)}, \frac{\lambda_1}{\tau} \biggr),
\]
where $\mathcal{S}(x,a)\equiv(\llvert x\rrvert -a)_+\sgn(x)$ is the
soft-thresholding function. We can write $\bOmega^*$ in a matrix form. Let
\begin{eqnarray}
\label{alg-D} \bD&=& 4(1+\gamma) (\bSigma_1\bOmega
\bSigma_1+\kappa\bSigma_2\bOmega\bSigma_2) +
2\gamma\bigl[ \tr(\bOmega\bSigma_1)\bSigma_1 +\kappa
\tr(\bOmega\bSigma_2)\bSigma_2 \bigr]\hspace*{-20pt}
\nonumber\\[-8pt]\\[-8pt]\nonumber
&&{} + 4\sym\bigl( \bSigma_1(\bOmega\bmu_1-\bdelta)
\bmu_1^{\top} + \kappa\bSigma_2(\bOmega
\bmu_2-\bdelta)\bmu_2^{\top} \bigr),
\nonumber
\end{eqnarray}
where $\sym(\bB)=(\bB+\bB^{\top})/2$ for any square matrix $\bB$.
By direct calculation,
%
\[
\bOmega^* = \mathcal{S} \biggl( \bOmega_0 - \frac{1}{\tau}\bD,
\frac{\lambda_1}{\tau} \biggr).
\]
%

We now describe our algorithm. Let us initialize $\bOmega^{(0)}=\bzero
_{d\times d}$, $\bdelta^{(0)}=\bzero$ and $\nu^{(0)}=0$. At
iteration $k$, the algorithm updates as follows:
\begin{itemize}
\item Compute $\bA=\bA(\bOmega^{(k-1)}, \bdelta^{(k-1)}, \nu
^{(k-1)})$ and $\mathbf{b}= \mathbf{b}(\bOmega^{(k-1)}, \bdelta^{(k-1)},
\nu^{(k-1)})$ using (\ref{alg-A,b}). Update $\bdelta^{(k)}=\argmin
_{\bdelta}\{\bdelta^{\top}\bA\bdelta-2\bdelta^{\top}\mathbf{b}+
\lambda_2\llvert \bdelta\rrvert _1\}$.
\item Compute\vspace*{2pt} $\bD=\bD(\bOmega^{(k-1)}, \bdelta^{(k)}, \nu
^{(k-1)})$ using (\ref{alg-D}). Update $\bOmega^{(k)} = \mathcal
{S}(\bOmega^{(k-1)} - \frac{1}{\tau}\bD, \frac{\lambda_1}{\tau
} )$.
\item Update $\nu^{(k)} = \nu^{(k-1)} + \rho[M(\bOmega
^{(k)},\bdelta^{(k)})-1]$.
\end{itemize}
Stop until $\max\{\rho\llvert \bOmega^{(k)}-\bOmega^{(k-1)}\rrvert,
\rho
\llvert \bdelta^{(k)}-\bdelta^{(k-1)}\rrvert, \llvert \nu^{(k)}-\nu
^{(k-1)}\rrvert /\rho\}
\leq\varepsilon$ for some pre-specified precision $\varepsilon$.

This is a modified version of the
augmented Lagrangian method, where in the step of updating $\bOmega$,
we minimize an upper envelope, which is obtained by locally linearizing
the augmented Lagrangian.

\begin{rem*}
QUADRO can be extended to folded concave penalties, for
example, to SCAD [\citet{fan2001variable}] or to adaptive Lasso
[\citet{zou2006adaptive}]. Using the Local Linear Approximation algorithm
[\citet{zou2008one,fan2014strong}], we can solve the SCAD-penalized
QUADRO and the adaptive-Lasso-penalized QUADRO by solving
$L_1$-penalized QUADRO with multiple-step and one-step iterations, respectively.
\end{rem*}

\section{Estimation of mean and covariance matrix} \label{secestimation}

QUADRO requires estimates of the mean vector and covariance matrix for
each class as inputs. We will show in Section~\ref{secanalysis} that
the performance of QUADRO is closely related to the max-norm estimation
error on mean vectors and covariance matrices.
Sample mean and sample covariance matrix work well for Gaussian data.
However, when data are from elliptical distributions, they may have
inferior performance as we estimate nonpolynomially many of means and
variances. In Sections~\ref{subsecmean}--\ref{subseccov}, we
suggest a robust $M$-estimator to estimate the mean and a rank-based
estimator to estimate the covariance matrix, which are more appropriate
for non-Gaussian data.
Moreover, in Section~\ref{subsecgamma} we discuss how to estimate the
model parameter $\gamma$ when it is unknown.

\subsection{Estimation of the mean} \label{subsecmean}

Suppose $\bx_1,\ldots,\bx_n$ are i.i.d. samples of a random vector
$\Xb=(X_1,\ldots, X_d)^{\top}$ from an elliptical distribution
$\mathcal{E}(\bmu, \bSigma, g)$. Let us denote $\bmu=(\mu_1,\ldots
,\mu_d)^{\top}$ and $\bx_i=(x_{i1},\ldots,x_{id})^{\top}$ for
$i=1,\ldots,n$. We estimate each $\mu_j$ marginally using the data
$\{ x_{1j}, \ldots, x_{nj}\}$.

One possible estimator is the sample median
\[
\widehat{\mu}_{Mj} = \operatorname{median} \bigl(\{x_{1j},
\ldots, x_{nj}\} \bigr).
\]
It can be shown that even under heavy-tailed distributions, $P (
\llvert \widehat{\mu}_{Mj}-\mu_j\rrvert > A\sqrt{\log(\delta^{-1})/n}
)\leq
\delta$ for small $\delta\in(0,1)$, where $A$ is a constant
determined by the probability density at $\mu_j$, for each fixed $j$.
This combined with the union bound gives that $\llvert \hbmu_M-\bmu
\rrvert _\infty
=O_p(\sqrt{\log(d)/n})$.

\citet{Catoni11} proposed another $M$-estimator for the mean of
heavy-tailed distributions. It works for distributions where mean is
not necessarily equal to median, which is essential for estimating
covariance of random variables. We denote the diagonal elements of the
covariance matrix $\bSigma$ as $\sigma^2_{1},\sigma^2_{2},\ldots
,\sigma^2_{d}$, and the off-diagonal elements as $\sigma_{kj}$ for
$k\neq j$.
The estimator $\widehat{\bmu}_C=(\widehat{\mu}_{C,1},\ldots,
\widehat{\mu}_{C,d})^{\top}$ is obtained as follows. For a strictly
increasing function $h\dvtx \mathbb{R}\rightarrow\mathbb{R}$ such that $
-\log(1-y+y^2/2)\leq h(y)\leq\log(1+y+y^2/2)$, and a value $\delta
\in(0,1)$ such that $n > 2 \log(1/\delta)$, we let
\[
\alpha_{\delta} = \biggl\{\frac{2\log(\delta^{-1})}{n [v +
\vafrac{2v\log(\delta^{-1})}{n-2\log(\delta^{-1})} ]} \biggr\}^{1/2},
\]
where $v$ is an upper bound of $\max\{\sigma_1^2,\ldots,\sigma_d^2\}
$. For each $j$, we define $\widehat{\mu}_{Cj}$ as the unique value
that satisfies
$\sum_{i=1}^n h(\alpha_\delta(x_{ij}-\widehat{\mu}_{Cj})) = 0$.
It was shown in \citet{Catoni11} that $P ( \llvert \widehat{\mu
}_{Cj}-\mu_j\rrvert > \sqrt{\tfrac{2v\log(\delta^{-1})}{n(1-2\log
(\delta
^{-1})/n)}} )\leq\delta$ when the variance of $X_j$ exists.
Therefore, by taking $\delta=1/(n\vee d)^2$, $\llvert \hbmu_M-\bmu
\rrvert _\infty
\leq C\sqrt{\log(d)/n}$ with probability at least $1-(n\vee d)^{-1}$,
which gives the desired convergence rate.

To\vspace*{1pt} implement this estimator, we take $h(y)= \operatorname{sgn}(y) \log
(1+\llvert y\rrvert +y^2/2)$.
For the choice of $v$, any value larger than $\max\{\sigma_1^2,
\ldots, \sigma_d^2\}$ would work in theory. \citet{Catoni11}
introduced a Lepski's adaptation method to choose $v$. For simplicity,
we take $v=3\max\{\widetilde{\sigma}_1^2,\ldots, \widetilde{\sigma
}^2_d\}$, where $\widetilde{\sigma}^2_j$ is the sample covariance of $X_j$.

The two estimators, the median and the $M$-estimator, both have a
convergence rate of $O_p(\sqrt{\log(d)/n})$ in terms of the max-norm
error. In our numerical experiments, the $M$-estimator has a better
numerical performance, and we stick to this estimator.

\subsection{Estimation of the covariance matrix} \label{subseccov}

To estimate the covariance matrix~$\bSigma$, we estimate the marginal
covariances $\{\sigma^2_j, 1\leq j\leq d\}$ and the correlation matrix
$\bC$ separately. Again, we need robust estimates even though the data
have fourth moments, as we simultaneously estimate nonpolynomial number
of covariance parameters.

First, we consider estimating $\sigma^2_{j}$. Note that $\sigma
_j^2=\mathbb{E}(X_j^2)-\mathbb{E}^2(X_j)$. We estimate $\mathbb
{E}(X_j^2)$ and $\mathbb{E}(X_j)$ separately. To estimate $\mathbb
{E}(X_j^2)$, we use the $M$-estimator described above on the squared
data $\{x^2_{1j}, \ldots, x^2_{nj}\}$ and denote the estimator by~$\widehat{\eta}_{Cj}$. This works as $\mathbb{E}(X_j^4)$ is finite
for each $j$ in our setting; in addition, the $M$-estimator applies to
asymmetric distributions. We then define
\[
\widehat{\sigma}_{Cj}^2 = \max\bigl\{\widehat{
\eta}_{Cj} - \widehat{\mu}_{Cj}^2,
\delta_0 \bigr\},
\]
where $\widehat{\mu}_{Cj}$ is the $M$-estimator of $\mathbb{E}(X_j)$
and $\delta_0>0$ is a small constant ($\delta_0<\min\{\sigma
_1^2,\ldots,\sigma_d^2\}$).
It is easy to see that when the fourth moments of $X_j$ are uniformly
upper bounded by a constant and $n\geq4\log(d^2)$, $\max\{\llvert
\widehat
{\sigma}_{Cj}-\sigma_j\rrvert, 1\leq j\leq d\}= O_p(\sqrt{\log(d)/n})$.

Next, we consider estimating the correlation matrix $\bC$. For this,
we use Kendall's tau correlation matrix proposed by \citet{HanLiu12}.
Kendall's tau correlation coefficients [\citet{Kendall}] are
defined as
\[
\tau_{jk} = \mathbb{P} \bigl( (X_j-\widetilde{X}_j)
(X_k-\widetilde{X}_k)> 0 \bigr) - \mathbb{P} \bigl(
(X_j-\widetilde{X}_j) (X_k-
\widetilde{X}_k)< 0 \bigr),
\]
where $\widetilde{\Xb}$ is an independent copy of $\Xb$. They have
the following relationship to the true coefficients:  $C_{jk}= \sin
(\frac{\pi}{2}\tau_{jk})$ for the elliptical family. Based on this
equality, we first estimate Kendall's tau correlation coefficients
using rank-based estimators
\[
\widehat{\tau}_{jk} = \cases{
\displaystyle \frac{2}{n(n-1)}\sum_{1\leq i< i'\leq n}\sgn\bigl((x_{ij}-x_{i'j})
(x_{ik}-x_{i'k}) \bigr), &\quad $j\neq k$,
\vspace*{3pt}\cr
1, &\quad $j=k$,}
\]
and then estimate the correlation matrix by $\widehat{\bC}=(\widehat
{C}_{jk})$ with
\[
\widehat{C}_{jk} = \sin\biggl(\frac{\pi}{2}\widehat{\tau
}_{jk} \biggr).
\]
It\vspace*{1pt} is shown in \citet{HanLiu12} that $\llvert \widehat{\bC}-\bC\rrvert
_\infty=
O_p(\sqrt{\log(d)/n})$.

Finally, we combine $\{\widehat{\sigma}^2_j, 1\leq j\leq d\}$ and
$\widehat{\bC}$ to get $\hbSigma$. Let
\[
\widetilde{\Sigma}_{jk} = \widehat{\sigma}_j \widehat{
\sigma}_k \widehat{C}_{jk}, \qquad1\leq j,k\leq d.
\]
It follows immediately that $\llvert \widetilde{\bSigma}-\bSigma
\rrvert _\infty=
O_p(\sqrt{\log(d)/n})$. However, this estimator is not necessarily
positive semi-definite. To implement QUADRO, we need $\widehat{\bSigma
}$ to be positive semi-definite so that the optimization in (\ref
{progquadro}) is a convex problem. We obtain $\hbSigma$ by projecting
$\widetilde{\bSigma}$ onto the cone of positive semi-definite
matrices through the convex optimization
%
%
\begin{equation}
\label{project} \hbSigma= \argmin_{\bA\dvtx  \bA~\mathrm
{is~positive~semidefinite}} \bigl\{ \llvert\bA- \widetilde{
\bSigma} \rrvert_\infty\bigr\}.
\end{equation}
Note\vspace*{1pt} that $\llvert \hbSigma-\widetilde{\bSigma}\rrvert _\infty\leq
\llvert \bSigma
-\widetilde{\bSigma}\rrvert _\infty$ by definition. Therefore,
$\llvert \hbSigma
-\bSigma\rrvert _\infty\leq\llvert \hbSigma- \widetilde{\bSigma
}\rrvert _\infty+
\llvert \widetilde{\bSigma}-\bSigma\rrvert _\infty\leq2\llvert
\widetilde{\bSigma
}-\bSigma\rrvert _\infty= O_p(\sqrt{\log(d)/n})$. To compute $\widehat
{\bSigma}$, we note that the optimization problem in (\ref{project})
can be formulated as the dual of a graphical lasso problem
corresponding to the smallest possible tuning parameter
that still guarantees a feasible solution [\citet{liu2012high}].
\citet{zhao2013psd} provide more algorithmic details.

\subsection{Estimation of kurtosis parameter} \label{subsecgamma}

When the kurtosis parameter $\gamma$ is unknown, we can estimate it
from data. Recall that $\gamma= \frac{1}{d(d+2)}\mathbb{E}(\xi^4)
-1$. Using decomposition (\ref{Zdecomposition}) and the properties of
$\Ub$, we have
\[
\mathbb{E} \bigl(\xi^4 \bigr) = \mathbb{E} \bigl\{ \bigl[(\Xb-
\bmu)^{\top}\bSigma^{-1}(\Xb-\bmu) \bigr]^2 \bigr
\}.
\]
Motivated by this equality, we propose the estimator
\[
\widehat{\gamma} = \max\Biggl\{ \frac{1}{d(d+2)} \frac{1}{n}\sum
_{i=1}^n \bigl[(\bx_i-
\widetilde{\bmu})^{\top}\widetilde{\bOmega}(\bx_i-
\widetilde{\bmu}) \bigr]^2 - 1, 0 \Biggr\},
\]
where $\widetilde{\bmu}$ and $\widetilde{\bOmega}$ are estimators
of $\bmu$ and $\bSigma^{-1}$, respectively.
\citet{kurtosis} considered a similar estimator in low-dimensional
settings, where they used the sample mean and sample covariance matrix.
In high dimensions, we a robust estimate to guarantee uniform
convergence. In particular, we take $\widetilde{\bmu}=\widehat{\bmu
}_C$ and $\widetilde{\bOmega} = \widehat{\bOmega}_{\mathrm
{clime}}$ where $\widehat{\bOmega}_{\mathrm{clime}}$ is the CLIME
estimator proposed in \citet{Clime}. We can also take the covariance
estimator in Section~\ref{subseccov}, but we will then need to
establish its sampling property as a precision matrix estimator. We
decide to use the CLIME estimator since such a property has already
been established by \citet{Clime}. Denote by $\bSigma^{-1}=(\Omega
_{jk})_{d\times d}$. From simple algebra,
\begin{eqnarray*}
\llvert\widehat{\gamma}-\gamma\rrvert&\leq&\max_{1\leq j,k\leq d}
\llvert
\widetilde{\mu}_j\widetilde{\Omega}_{jk}\widetilde{
\mu}_k - \mu_j\Omega_{jk}\mu_k
\rrvert
\\
&\leq& C\max\bigl\{ \llvert\widetilde{\bmu}-\bmu\rrvert
_{\infty},
\bigl\llvert\widetilde{\bOmega}-\bSigma^{-1} \bigr\rrvert
_{\infty} \bigr\}.
\end{eqnarray*}
In\vspace*{1pt} Section~\ref{subsecmean}, we have seen that $\Vert\widehat{\bmu
}_C - \bmu\Vert_\infty=O_p(\sqrt{\log(d)/n})$. Moreover, \citet
{Clime} showed that $\llvert \widetilde{\bOmega}- \bSigma^{-1}\rrvert
_\infty=
\Vert\bSigma^{-1}\Vert_1 \cdot O_p(\sqrt{\log(d)/n})$ under mild
conditions, where $\Vert\cdot\Vert_1$ is the matrix $L_1$-norm.
Therefore, provided that $\Vert\bSigma^{-1}\Vert_1\leq C$, we
immediately have $\llvert \widehat{\gamma}-\gamma\rrvert =O_p(\sqrt
{\log(d)/n})$.


\section{Theoretical properties} \label{secanalysis}

In this section, we establish an oracle inequality for the Rayleigh
quotient of the QUADRO estimates $(\hbOmega,\hbdelta)$.
We assume that $\pi$ and $\gamma$ are known. For notational
simplicity, we set $\lambda_1=\lambda_2=\lambda$. The results can be
easily generalized to the case $\lambda_1\neq\lambda_2$. Moreover,\vspace*{1pt}
we drop the symmetry constraint $\bOmega=\bOmega^{\top}$ in all
optimization problems involved. This simplifies the expression of the
regularity conditions. The analysis with the symmetry constraint is a
trivial extension of current analysis.

Recall the definition of $M$, $L_1$ and $L_2$ in (\ref{ML1L2}) and
$\kappa=(1-\pi)/\pi$ and $L=L_1+\kappa L_2$, the Rayleigh quotient
of $(\bOmega, \bdelta)$ is equal to (up to a multiplicative constant)
\[
R(\bOmega, \bdelta) = \frac{[M(\bOmega,\bdelta)]^2}{L(\bOmega,
\bdelta)}.
\]
The QUADRO estimates are
\[
(\hbOmega,\hbdelta) = \argmin_{(\bOmega,\bdelta)\dvtx  \widehat
{M}(\bOmega,\bdelta)=1} \bigl\{ \widehat{L}(\bOmega,\bdelta
)+\lambda\llvert\bOmega\rrvert_1 + \lambda\llvert\bdelta\rrvert
_1 \bigr\}.
\]



We shall compare the Rayleigh quotient of $(\hbOmega,\hbdelta)$ with
the Rayleigh quotients of a class of ``oracle solutions.'' This class
includes the one that maximizes the true Rayleigh quotient, which we
denote by $(\bOmega^*_0, \bdelta^*_0)$. Here we adopt a class of
solutions as the ``oracle'' instead of only $(\bOmega^*_0,\bdelta
^*_0)$, because we want the results not tied to the sparsity assumption
on $(\bOmega^*_0, \bdelta^*_0)$ but a weaker assumption: at least one
solution in this class is sparse.

Our theoretical development is technically nontrivial. Conventional
oracle inequalities are derived in a setting of minimizing a
data-dependent loss without constraint, and the risk function is the
expectation of the loss. Here we minimize a data-dependent loss with a
data-dependent equality constraint, and the risk function---the
Rayleigh quotient---is not equal to the expectation of the loss.
A~similar setting was considered in \citet{ROAD}, where they introduced
a data-dependent intermediate solution to deal with such equality
constraint. However, the rate they obtained depends on this
intermediate solution, which is very hard to quantify. In contrast, the
rate in our results purely depends on the oracle solution.
To get rid of the intermediate solution in the rate, we need to
carefully quantify its difference from both the QUADRO solution and the
oracle solution. The technique is new, and potentially useful for other
problems.



\subsection{Oracle solutions, the restricted eigenvalue condition}

For any $\lambda_0\geq0$, we define the \emph{oracle solution
associated with $\lambda_0$} to be
%
%
\begin{equation}
\label{oracle} \bigl(\bOmega^*_{\lambda_0}, \bdelta^*_{\lambda_0} \bigr
) =
\argmin_{(\bOmega
,\bdelta)\dvtx  M(\bOmega,\bdelta)=1} \bigl\{ L(\bOmega,\bdelta)+\lambda
_0\llvert
\bOmega\rrvert_1 + \lambda_0\llvert\bdelta\rrvert
_1 \bigr\}.
\end{equation}
We shall compare the Rayleigh quotient of $(\hbOmega, \hbdelta)$ to
that of $(\bOmega^*_{\lambda_0}, \bdelta^*_{\lambda_0})$, for an
arbitrary $\lambda_0$.
In particular, when $\lambda_0=0$, the associated oracle solution (may
not be unique) becomes
\[
\bigl(\bOmega^*_0,\bdelta^*_0 \bigr) =
\argmin_{(\bOmega,\bdelta)\dvtx  M(\bOmega
,\bdelta)=1} \bigl\{ L(\bOmega,\bdelta) \bigr\}.
\]
It maximizes the true Rayleigh quotient.

Next, we introduce a restricted eigenvalue (RE) condition jointly on
$\bSigma_1$, $\bSigma_2$, $\bmu_1$ and $\bmu_2$.
For any matrices $\bA$ and $\bB$, let $\vec(\bA)$ be the
vectorization of $\bA$ by stacking all the elements of $\bA$ column
by column, and $\bA\otimes\bB$ be the Kronecker product of $\bA$
and $\bB$. We define the matrices
\[
\bQ_k = %
\lleft[ \matrix{ \bigl(2(1+\gamma)
\bSigma_k+ 4\bmu_k\bmu_k^{\top}
\bigr)\otimes\bSigma_k + \gamma\vec(\bSigma_k)\vec(
\bSigma_k)^{\top} & - 4\bmu_k\otimes
\bSigma_k
\vspace*{3pt}\cr
- 4\bmu_k ^{\top}\otimes
\bSigma_k & 4\bSigma_k} \rright],
\]
for $k=1,2$.
We note that there are $(d^2+d)$ coefficients to decide when maximizing
$R(\bOmega, \bdelta)$:  $d^2$ elements of $\bOmega$ and $d$ elements
of $\bdelta$. We can stack all these coefficients into a long vector
$\bx=\bx(\bOmega,\bdelta)$ in $\mathbb{R}^{d^2+d}$ defined as
%
%
\begin{equation}
\label{xOmega,delta} \bx(\bOmega, \bdelta) \equiv%
\bigl[ \matrix{\vec(\bOmega)^{\top}, \bdelta^{\top}} \bigr]^{\top}.
\end{equation}
It can be shown that $L_k(\bOmega,\bdelta)=\bx^{\top}\bQ_k\bx$,
for $k=1,2$; see Lemma \ref{lemvectorize}. Therefore, $L(\bOmega
,\bdelta)=\bx^{\top}\bQ\bx$, where $\bQ=\bQ_1+\kappa\bQ_2$.
Our RE condition is then imposed on the $(d^2+d)\times(d^2+d)$ matrix
$\bQ$, and hence implicitly on $(\bSigma_1, \bSigma_2, \bmu_1, \bmu_2)$.

We now formally introduce the RE condition. For a set $S\subset\{1,2,
\ldots, d^2+d\}$ and a nonnegative value $\bar{c}$, we define the
\textit{restricted eigenvalue} in the following way:
\[
\Theta(S; \bar{c}) = \min_{\bv\dvtx \llvert \bv_{S^c}\rrvert _1\leq\bar{c}
\llvert \bv
_S\rrvert _1}\frac{\bv^{\top}\bQ\bv}{\llvert \bv_S\rrvert ^2}.
\]

Generally speaking, $\Theta(S; \bar{c})$ depends on $(\bSigma
_1,\bSigma_2, \bmu_1, \bmu_2)$ in a complicated way. For $\bar
{c}=0$, the following proposition builds a connection between $\Theta
(S;0)$ and $(\bSigma_1, \bSigma_2, \bmu_1,\bmu_2)$. For each
$S\subset\{1,2, \ldots, d^2+d\}$, there exist sets $U\subset\{
1,\ldots,d\}\times\{1,\ldots,d\}$ and $V\subset\{1,\ldots,d\}$
such that the support of $\bx(\bOmega,\bdelta)$ is $S$ if and only
if the support of $\bOmega$ is $U$ and the support of $\bdelta$ is
$V$. Let
\[
U'= \bigcup_{(i,j)\in U}\{i,j\}.
\]
Then $U\subset U'\times U'$. The following result is proved in \citet
{QUADROsupp}.

%
%
\begin{prop} \label{propQeigen}
For any set $S\subset\{1,\ldots,d^2+d\}$, suppose $U'$ and $V$ are
defined as above.
Let $\widetilde{\bSigma}_k$ be the submatrix of $\bSigma_k$ by
restricting rows and columns to $U'\cup V$, $\widetilde{\bmu}_k$ be
the subvector of $\bmu_k$ by constraining elements to $U'\cup V$, for $k=1,2$.
If there exist constants $v_1, v_2>0$ such that $\lambda_{\min} (
\widetilde{\bSigma}_{k} - v_1 \widetilde{\bmu}_k\widetilde{\bmu
}_k^{\top} )\geq\frac{1}{2}\lambda_{\min}(\widetilde{\bSigma
}_k)\geq\frac{v_2}{2}$ for $k=1,2$, then
\[
\Theta(S, 0) \geq(1+\gamma) (1+\kappa)v_2 \min\biggl\{
v_2, \frac
{4v_1}{2+v_1(1+\gamma)} \biggr\} >0.
\]
\end{prop}
%

\subsection{Oracle inequality on Rayleigh's quotient} \label{subsecoracle}

Suppose $\max\{\llvert \bSigma_k\rrvert _\infty, \break \llvert \bmu
_k\rrvert _\infty, k=1,2\} \leq
1$ and $\llvert \hbSigma_k-\bSigma_k\rrvert _\infty\leq\llvert
\bSigma_k\rrvert _\infty$,
$\llvert \hbmu_k-\bmu_k\rrvert _\infty\leq\llvert \bmu_k\rrvert
_\infty$ for $k=1,2$, without
loss of generality. For any $\lambda_0\geq0$, let $(\bOmega
^*_{\lambda_0}, \bdelta^*_{\lambda_0})$ be the\vspace*{1pt} associated oracle
solution and $S$ be the\vspace*{1pt} support of $\bx^*_{\lambda_0}=[\vec(\bOmega
^*_{\lambda_0})^{\top}, (\bdelta^*_{\lambda_0})^{\top}]^{\top}$.
Let $\Delta_n=\max\{\llvert \hbSigma_k-\bSigma_k\rrvert _\infty,
\llvert \hbmu_k-\bmu
_k\rrvert _\infty, k = 1,2\}$. We have the following result for any given
estimators, the proof of which we postpone to Section~\ref{secproof}.

%
%
\begin{teo} \label{teoRbound}
Given $\lambda_0\geq0$, let $S$ be the support of $\bx^*_{\lambda
_0}$, $s_0=\llvert S\rrvert $ and $k_0=\max\{s_0, R(\bOmega^*_{\lambda
_0}, \bdelta
^*_{\lambda_0})\}$.
Suppose that $\Theta(S,0)\geq c_0$, $\Theta(S, 3)\geq a_0$ and
$R(\bOmega^*_{\lambda_0}, \bdelta^*_{\lambda_0})\geq u_0$, for some
positive constants $a_0$, $c_0$ and $u_0$. We assume $4s_0\Delta
_n^2\leq a_0c_0$ and $\max\{s_0\Delta_n, s_0^{1/2}k_0^{1/2}\lambda
_0\}<1$ without loss of generality. Then there exist positive constants
$C=C(a_0,c_0, u_0)$ and $A=A(a_0, c_0, u_0)$ such that for any $\eta>1$,
\[
\frac{R(\hbOmega,\hbdelta)}{R(\bOmega^*_{\lambda_0},\bdelta
^*_{\lambda_0})} \geq1 - A \eta^2 \max\bigl\{s_0
\Delta_n, s_0^{1/2}k_0^{1/2}
\lambda_0 \bigr\},
\]
by taking $\lambda= C\eta\max\{ s_0^{1/2}\Delta_n, k_0^{1/2}\lambda
_0\}[R(\bOmega^*_{\lambda_0},\bdelta^*_{\lambda_0})]^{-1/2}$.
\end{teo}

In Theorem \ref{teoRbound}, the rate of convergence has two parts.
The term $s_0\Delta_n$ reflects how the stochastic errors of
estimating $(\bSigma_1, \bSigma_2, \bmu_1, \bmu_2)$ affect the
Rayleigh quotient. The term $s_0^{1/2}k_0^{1/2}\lambda_0$ is an extra
term that depends on the oracle solution we aim to use for comparison.
In particular, if we compare $R(\widehat{\bOmega},\widehat{\bdelta
})$ with $R_{\max} \equiv R(\bOmega^*_0, \bdelta^*_0)$, the
population maximum Rayleigh quotient with $\lambda_0 = 0$, this extra
term disappears. If we further use the estimators in Section~\ref
{secestimation}, $\Delta_n=O_p(\sqrt{\log(d)/n})$.
We summarize the result as follows.

%
%
\begin{coro} \label{coroquadrorate1}
Suppose that the condition of Theorem~\ref{teoRbound} holds with
\mbox{$\lambda_0 = 0$}. Then for some positive constants $A$ and $C$, when
$\lambda> Cs_0^{1/2}R_{\max}^{-1/2}\Delta_n$, we have
\[
R(\hbOmega,\hbdelta)\geq(1 - A s_0\Delta_n
)R_{\max}.
\]
Furthermore, if the mean vectors and covariance matrices are estimated
by using the robust methods in Section~\ref{secestimation}, then when
$\lambda> C s_0^{1/2}R_{\max}^{-1/2}\sqrt{\log(d)/n}$,
\[
R(\hbOmega,\hbdelta)\geq\bigl( 1 - A s_0 \sqrt{\log(d)/n}
\bigr)R_{\max},
\]
with probability at least $1-(n\vee d)^{-1}$.
\end{coro}

From Corollary \ref{coroquadrorate1}, when $(\bOmega^*_0, \bdelta
^*_0)$ is truly sparse, $R(\hbOmega, \hbdelta)$ is close to the
population maximum Rayleigh quotient $R_{\max}$. However, we note that
Theorem \ref{teoRbound} considers more general situations, including
cases where $(\bOmega^*_0, \bdelta^*_0)$ is not sparse.
As long as there exists an ``approximately optimal'' and sparse
solution, that is, for a small $\lambda_0$ the associated oracle
solution $(\bOmega^*_{\lambda_0}, \bdelta^*_{\lambda_0})$ is
sparse, Theorem \ref{teoRbound} guarantees that $R(\hbOmega,
\hbdelta)$ is close to $R(\bOmega^*_{\lambda_0}, \bdelta^*_0)$ and
hence close to $R_{\max}$.




\begin{rem*}
Our results are analogous to oracle inequalities for
prediction error in linear regressions; therefore, the condition
$\Theta(S,\bar{c})$ is similar to the RE condition in linear
regressions [\citet{bickel2009simultaneous}]. To recover the
support of
$(\bOmega^*_0,\bdelta^*_0)$, conditions similar to the
``irrepresentable condition'' for Lasso [\citet{zhao2006model}]
are needed.
\end{rem*}

\section{Application to classification} \label{secclassification}

One important application of QUADRO is high-dimensional classification
for elliptically-distributed data. Suppose $(\hbOmega, \hbdelta)$ are
the QUADRO estimates. This yields the classification rule
\[
\widehat{h}(\bx) = I \bigl\{ \bx^{\top}\hbOmega\bx-2\hbdelta
^{\top}\bx< c \bigr\}.
\]
In this section, we first show that for normally distributed data, the
Rayleigh quotient is a proxy of the classification error, and then
derive an analytic choice of $c$. Comparing with many other
high-dimensional classification methods, QUADRO produces quadratic
boundaries and can handle both non-Gaussian distributions and nonequal
covariance matrices.


\subsection{Approximation of classification errors} \label{subsecerrapprox}

Given $(\bOmega, \bdelta)$ and a threshold $c$, a~general quadratic
rule $h(\bx)=h(\bx; \bOmega,\bdelta,c)$ is defined as
%
%
\begin{equation}
\label{quadrule} h(\bx; \bOmega,\bdelta,c)= I \bigl\{ \bx^{\top}\bOmega
\bx- 2\bx^{\top}\bdelta< c \bigr\}.
\end{equation}
%
We reparametrize $c$ as
%
%
\begin{equation}
\label{ct} c= t M_1(\bOmega,\bdelta)+(1-t)M_2(
\bOmega,\bdelta).
\end{equation}
Here $M_k(\bOmega, \bdelta) = \bmu^{\top}_k\bOmega\bmu_k - 2\bmu
_k^{\top} \bdelta+ \tr(\bOmega\bSigma_k)$ is the mean of $Q(\Xb)$
in class $k$, for $k=1,2$.
After the reparametrization, $t$ is \emph{scale-free}. As we will see
below, in most cases, given $\bOmega$ and $\bdelta$, the optimal $t$
that minimizes the classification error takes values on $(0,1)$.

From now on, we write $h(\bx; \bOmega,\bdelta, c)=h(\bx; \bOmega,
\bdelta, t)$. Let $\Err(\bOmega, \bdelta, t)$ be the classification
error of $h(\cdot; \bOmega,\bdelta,t)$. Due to technical
difficulties, we only give results for Gaussian distributions. Suppose
$\Xb\mid(Y=0)\sim\mathcal{N}(\bmu_1,\bSigma_1)$ and $\Xb\mid(Y=1)\sim
\mathcal{N}(\bmu_2,\bSigma_2)$.
For $k=1,2$, we write
\[
\bSigma_k^{1/2}\bOmega\bSigma_k^{1/2}=
\bK_k\bS_k\bK_k^T,
\]
where $\bS_k$ is a diagonal matrix containing the nonzero eigenvalues,
and the columns of $\bK_k$ are corresponding eigenvectors. Let $\bbeta
_k=\bK_k^T\bSigma_k(\bOmega\bmu_k-\bdelta)$. When $\max\{\llvert \bS
_k\rrvert _\infty, \llvert \bbeta_k\rrvert _\infty, k=1,2\}$ is bounded,
the following proposition shows that an approximation of $\Err(\bOmega
, \bdelta, t)$ is
\[
\overline{\Err}(\bOmega, \bdelta, t) \equiv\pi\bar{\Phi} \biggl(
\frac{(1-t)M(\bOmega,\bdelta)}{\sqrt{L_1(\bOmega,\bdelta)}} \biggr) +
(1-\pi)\bar{\Phi} \biggl( \frac{tM(\bOmega,\bdelta)}{\sqrt
{L_2(\bOmega,\bdelta)}} \biggr),
\]
where $M$, $L_1$ and $L_2$ are defined in (\ref{ML1L2}), $\Phi$ is
the distribution function of a standard normal variable and $\bar{\Phi
}=1-\Phi$. Its proof is contained in Section~\ref{secproof}.


%
%
\begin{prop} \label{properrgap}
Suppose that $\max\{\llvert \bS_k\rrvert _\infty, \llvert \bbeta
_k\rrvert _\infty, k=1,2\}
\leq C_0$ for some constant $C_0>0$, and let $q$ be the rank of
$\bOmega$. Then as $d$ goes to infinity,
\[
\bigl\llvert\Err(\bOmega, \bdelta, t) - \overline{\Err}(\bOmega,
\bdelta, t)
\bigr\rrvert= \frac{O(q) + o(d) }{[\min\{L_1(\bOmega,\bdelta
), L_2(\bOmega,\bdelta)\}]^{3/2}}.
\]
\end{prop}

In particular, if we consider all such $(\bOmega,\bdelta)$
that the variance of $Q(\Xb;\bOmega,\bdelta)$ under both classes are
lower bounded by $c_0d^{\theta}$ for some constants $\theta> 2/3$ and
$c_0>0$, then we have $\llvert \Err-\overline{\Err}\rrvert =o(1)$.

%
%
\begin{figure}[b]

\includegraphics{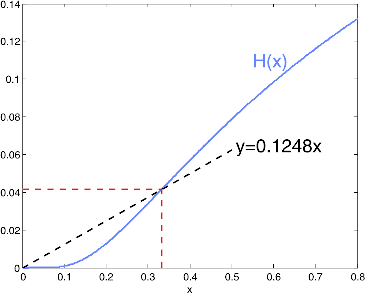}

\caption{Function $H(x)=\bar{\Phi}(1/\sqrt{x})$.} \label{figH}
\end{figure}

We now take a closer look at $\overline{\Err}$. Let $H(x)=\bar{\Phi
}(1/\sqrt{x})$, which is monotone increasing on $(0, \infty)$.
Writing for short $M=M_1-M_2$, $M_k=M_k(\bOmega,\bdelta)$ and
$L_k=L_k(\bOmega,\bdelta)$ for $k=1,2$, we have
\[
\overline{\Err}(\bOmega, \bdelta, t) = \pi H \biggl( \frac
{L_1}{(1-t)^2 M^2} \biggr)
+ (1-\pi) H \biggl( \frac{L_2}{t^2 M^2} \biggr).
\]
Figure~\ref{figH} shows that $H(\cdot)$ is nearly linear on an
important range. This suggests the following approximation:
%
%
\begin{equation}
\label{Happrox} \qquad\overline{\Err}(\bOmega, \bdelta, t) \approx H \biggl(
\pi
\frac
{L_1}{(1-t)^2 M^2} + (1-\pi) \frac{L_2}{t^2 M^2} \biggr) = H \biggl(
\frac{\pi}{(1-t)^2}\frac{1}{R^{(t)}} \biggr),
\end{equation}
where $R^{(t)}=R^{(t)}(\bOmega,\bdelta)$ is the $R(\bOmega,\bdelta
)$ in (\ref{defR}) corresponding to the $\kappa$ value
\[
\kappa(t)\equiv\frac{1-\pi}{\pi}\frac{(1-t)^2}{t^2}.
\]
The approximation in (\ref{Happrox}) is quantified in the following
proposition, which is proved in \citet{QUADROsupp}. 
%

\begin{prop} \label{propH-Taylor}
Given $(\bOmega, \bdelta, t)$, we write for short $R_k= R_k(\bOmega
,\bdelta)=[M(\bOmega, \bdelta)]^2/L_k(\bOmega, \bdelta)$, for
$k=1,2$, and define
\begin{eqnarray*}
V_1 & =& V_1(\bOmega, \bdelta, t)= \min\biggl
\{(1-t)^2 R_1, \frac
{1}{(1-t)^2 R_1} \biggr\},
\\
V_2 & =& V_2(\bOmega, \bdelta, t)= \min\biggl
\{t^2 R_2, \frac
{1}{t^2 R_2} \biggr\},
\\
V &=& V(
\bOmega, \bdelta, t) = \max\{ V_1/V_2,
V_2/V_1 \}.
\end{eqnarray*}
Then there exists a constant $C>0$ such that
\[
\biggl\llvert\overline{\Err}(\bOmega, \bdelta, t)- H \biggl(\frac{\pi
}{(1-t)^2R^{(t)}(\bOmega,\bdelta)}
\biggr) \biggr\rrvert\leq C \bigl[\max\{ V_1, V_2\}
\bigr]^{1/2} \cdot\llvert V-1\rrvert^2.
\]
In particular, when $t=1/2$,
\[
\biggl\llvert\overline{\Err}(\bOmega, \bdelta, t)- H \biggl(\frac{\pi
}{(1-t)^2R^{(t)}(\bOmega,\bdelta)}
\biggr) \biggr\rrvert\leq C {R_0^{1/2}}\cdot\biggl(
\frac{\Delta R}{R_0} \biggr)^{2},
\]
where $R_0=\max\{\min\{R_1,1/{R_1}\},\min\{R_2,1/{R_2}\}\}$ and
$\Delta R=\llvert R_1-R_2\rrvert $.
\end{prop}

Note that $L_1$ and $L_2$ are the variances of $Q(\Xb)=\Xb^{\top
}\bOmega\Xb- 2\Xb^{\top}\bdelta$ for two classes, respectively. In
cases where $\llvert L_1-L_2\rrvert \ll\min\{L_1, L_2\}$, $\Delta
R\ll R_0$. Also,
$R_0$ is always bounded by 1, and it tends to $0$ in many situations,
for example, when $R_1, R_2\to\infty$, or $R_1, R_2\to0$, or $R_1\to
0, R_2\to\infty$. Proposition \ref{propH-Taylor} then implies that
the approximation in (\ref{Happrox}) when $t=1/2$ is good.

Combining Propositions \ref{properrgap} and \ref{propH-Taylor}, the
classification error of a general quadratic rule $h(\cdot; \bOmega,
\bdelta, t)$ is approximately a monotone decreasing transform of the
Rayleigh quotient $R^{(t)}(\bOmega, \bdelta)$, corresponding to
$\kappa=\kappa(t)$. In particular, when $t=1/2$ [i.e.,
$c=(M_1+M_2)/2$], $R^{(1/2)}(\bOmega,\bdelta)$ is exactly the one
used in QUADRO. Consequently, if we fix the threshold to be
$c=(M_1+M_2)/2$, then the Rayleigh quotient (upon with a monotone
transform) is a good proxy for classification error. This explains why
Rayleigh-quotient based procedures can be used for classification.

\begin{rem*}
Even in the region that $H(\cdot)$ is far from being
linear such that the upper bound in Proposition \ref{propH-Taylor} is
not $o(1)$, we can still find a monotone transform of the Rayleigh
quotient as an \textit{upper bound} of the classification error. To see
this, note that for $x\in[1/3,\infty)$, $H(x)$ is a concave function.
Therefore, the approximation in (\ref{Happrox}) becomes an inequality,
that is, $\overline{\Err}(\bOmega,\bdelta,t)\leq H ( \tfrac
{\pi R^{(t)}}{(1-t)^2} )$. For $x\in(0, 1/3)$, $H(x)\leq
0.1248x$. It follows that $\overline{\Err}(\bOmega,\bdelta,t)\leq
0.1248\cdot\tfrac{\pi R^{(t)}}{(1-t)^2}$.
\end{rem*}

\begin{rem*}
In the current setting, the Bayes classifier is a
quadratic rule $h(\bx; \bOmega_B,\bdelta_B, c_B)$ with $\bOmega_B =
\bSigma_1^{-1}-\bSigma_2^{-1}$, $\bdelta_B = \bSigma_1^{-1}\bmu_1
- \bSigma_2^{-1}\bmu_2$ and $c_B = \bmu_2^{\top}\bSigma_2^{-1}\bmu
_2-\bmu_1^{\top}\bSigma_1^{-1}\bmu_1$. Let $(\bOmega^*_0, \bdelta
^*_0)$ be the population solution of QUADRO when $\lambda=0$. We note
that $(\bOmega_B, \bdelta_B)$ and $(\bOmega^*_0, \bdelta^*_0)$ are
different: the former minimizes $\inf_t\Err(\bOmega,\bdelta,t)$,
while the latter minimizes $\overline{\Err}(\bOmega,\bdelta,1/2)$.
\end{rem*}

\subsection{QUADRO as a classification method}

Results in Section~\ref{subsecerrapprox} suggest an analytic method
to choose the threshold $c$, or equivalently $t$, with given $(\bOmega
,\bdelta)$. Let
%
%
\begin{equation}
\label{hatt} \widehat{t} \in\min_t \biggl\{ \pi\bar{
\Phi} \biggl(\frac
{(1-t)\widehat{M}(\bOmega,\bdelta)}{\sqrt{\widehat{L}_1(\bOmega
,\bdelta)}} \biggr) + (1-\pi)\bar{\Phi} \biggl(
\frac{t\widehat{M}(\bOmega,\bdelta
)}{\sqrt{\widehat{L}_2(\bOmega,\bdelta)}} \biggr) \biggr\},
\end{equation}
and set
%
%
\begin{equation}
\label{hatc} \widehat{c}= (1-\widehat{t})\widehat{M}_1(\bOmega,
\bdelta) + \widehat{t}\widehat{M}_2(\bOmega,\bdelta).
\end{equation}
Here (\ref{hatt}) is a one-dimensional optimization problem and can be
solved easily.
The resulting QUADRO classification rule is
\[
\widehat{h}^{\mathrm{Quad}}(\bx) = I \bigl\{\bx^{\top}\hbOmega\bx- 2
\bx^{\top
}\hbdelta- \widehat{c}<0 \bigr\}.
\]


As a by-product, the method to decide $c$, described in (\ref{hatt})
and (\ref{hatc}), can be used in other classification procedures on
Gaussian data, such as logistic regression, quadratic\vspace*{1pt} discriminant
analysis (QDA) and kernel support vector machine, once $(\hbOmega
,\hbdelta)$ are given. It provides a fast and purely data-driven way
to decide the threshold value in quadratic classification rules. In our
numerical experiments, it performs well.

\section{Numerical studies} \label{secnumerical}

In this section, we investigate the performance of QUADRO in several
simulation examples and a real data example. The simulation studies
contain both Gaussian models and general elliptical models. We compare
QUADRO with several \textit{classification-oriented procedures}.
Performances are evaluated in terms of classification errors.

\subsection{Simulations under Gaussian models}

Let $n_1=n_2=50$ and $d=40$. For each given $\bmu_1$, $\bmu_2$,
$\bSigma_1$ and $\bSigma_2$, we generate $100$ training datasets
independently, each with $n_1$ data from
$\mathcal{N}(\bmu_1,\bSigma_1)$ and $n_2$ data from $\mathcal
{N}(\bmu_2,\bSigma_2)$.
In QUADRO, we input the sample means and sample covariance matrices. We
set $\lambda_2=r\lambda_1$ and work with $\lambda_1$ and $r$ from
now on. The two tuning parameters $\lambda_1\geq0$ and $r>0$ are
selected in the following way. For various pairs of $(\lambda_1, r)$,
we apply QUADRO for each pair and evaluate the classification error via
4000 newly generated testing data; we then choose the $(\lambda_1,
r)$ that minimize the classification error.

We compare QUADRO with five \textit{classification-oriented procedures}:
\begin{itemize}
\item Sparse logistic regression (SLR): We apply the sparse logistic
regression to the augmented feature space $\{X_i, 1\leq i\leq d;
X_iX_j, 1\leq i\leq j\leq d\}$. The resulting estimator then gives a
quadratic projection with $(\bOmega, \bdelta, c)$ decided from the
fitted regression coefficients. We implement the sparse logistic
regression using the R package {glmnet}.
\item Linear sparse logistic regression (L-SLR): We apply the sparse
logistic regression directly to the original feature space $\{X_i,
1\leq i\leq d\}$.
\item ROAD [\citet{ROAD}]: This is a linear classification
method, which
can be formulated equivalently as a modified version of QUADRO by
enforcing $\widehat{\Omega}$ as the zero matrix and plugging in the
pooled sample covariance matrix.
\item Penalized-LDA (P-LDA) [\citet{witten2011penalized}]: This
is a
variant of LDA, which solves an optimization problem with a nonconvex
objective and $L_1$ penalties. Also, P-LDA only uses diagonals of the
sample covariance matrices.
\item FAIR [\citet{FAIR}]: This is a variant of LDA for high-dimensional
settings, where screening is adopted to pre-select features and only
the diagonals of the sample covariance matrices are used.
\end{itemize}
To make a fair comparison, the tuning parameters in SLR and L-SLR are
selected in the same way as in QUADRO based on 4000 testing data. ROAD
and P-LDA are self-tuned by its package. The number of features chosen
in FAIR is calculated in the way suggested in [\citet{FAIR}].

\begin{figure}

\includegraphics{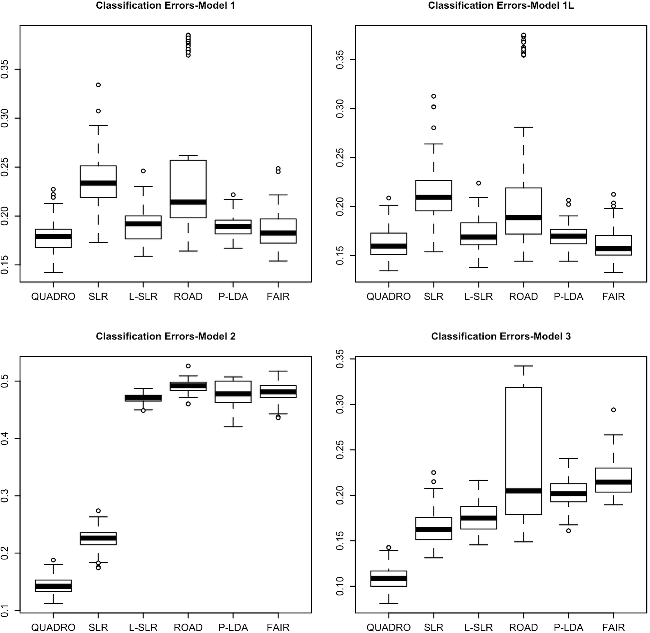}

\caption{Distributions of minimum classification error based on 100
replications for four different normal models. The tuning parameters
for QUADRO, SLR and L-SLR are chosen to minimize the classification
errors of 4000 testing samples. See \citet{QUADROsupp} for detailed
numerical tables.} \label{figClassiErrorGaussian}
\end{figure}

We consider four models:
\begin{itemize}[--]
\item[--] \textit{Model} 1: $\bSigma_1$ is the identity matrix. $\bSigma
_2$ is a diagonal matrix in which the first 10 elements are equal to
1.3 and the rest are equal to 1. $\bmu_1=\bzero$, and $\bmu
_2=(0.7,\ldots,0.7,0,\ldots,0)^{\top}$ with the first 10 elements of
$\bmu_2$ being nonzero.
\item[--] \textit{Model} 1L: $\bmu_1$, $\bmu_2$ are the same as in {model}~1, and both $\bSigma_1$ and $\bSigma_2$ are the identity matrix.
\item[--] \textit{Model} 2: $\bSigma_1$ is a block-diagonal matrix. Its
upper left $20\times20$ block is an equal correlation matrix with
$\rho=0.4$, and its\vspace*{1pt} lower right $20\times20$ block is an identity
matrix. $\bSigma_2=(\bSigma^{-1}_1+\,\bI)^{-1}$. We also set $\bmu
_1=\bmu_2=\bzero$. In this model, neither $\bSigma_1^{-1}$ nor
$\bSigma_2^{-1}$ is sparse, but $\bSigma_1^{-1}-\bSigma_2^{-1}$ is.
\item[--] \textit{Model} 3: $\bSigma_1$, $\bSigma_2$ and $\bmu_1$ are
the same as in {model~2}, and $\bmu_2$ is taken from {model~1}.
\end{itemize}

Figure~\ref{figClassiErrorGaussian} contains the boxplots for the
classification errors of all methods.
In all four models, QUADRO outperforms other methods in terms of
classification error.
In model~1L, $\bSigma_1=\bSigma_2$, so the Bayes classifier is
linear. In this case which favors linear methods, QUADRO is still
competitive with the best of all linear classifiers.
In model~2, $\bmu_1=\bmu_2$, so linear methods can do no better than
random guessing. Therefore, ROAD, L-SLR, P-LDA and FAIR all have very
poor performances. For the two quadratic methods, QUADRO is
significantly better than SLR.
In models 1~and~3, $\bmu_1\neq\bmu_2$ and $\bSigma_1\neq\bSigma
_2$, so in the Bayes classifier, both ``linear'' parts and
``quadratic'' parts play important roles. In model~1, both $\bSigma_1$
and $\bSigma_2$ are diagonal, and the setting favors methods using
only diagonals of sample covariance matrices. As a result, P-LDA and
FAIR perform quite well. In model~3, $\bSigma_1$ and $\bSigma_2$ are
both nondiagonal and nonsparse (but $\bSigma_1-\bSigma_2$ is
sparse). We see that the performances of P-LDA and FAIR are
unsatisfactory. QUADRO outperforms other methods in both models 1~and~3.

Comparing SLR and L-SLR, we see the former considers a broader class,
while the latter is more robust, but neither of them perform uniformly
better. However, QUADRO performs well in all cases. In terms of
Rayleigh quotients, QUADRO also outperforms other methods in most cases.

\subsection{Simulations under elliptical models}

Let $n_1=n_2=50$ and $d=40$. For each given $\bmu_1$, $\bmu_2$,
$\bSigma_1$ and $\bSigma_2$, data are generated from multivariate t
distribution with degrees of freedom $5$. In QUADRO, we input the
robust $M$-estimators for means and the rank-based estimators for
covariance matrices as described in Section~\ref{secestimation}. We
compare the performance of QUADRO with the five methods compared under
Gaussian settings. We also implement QUADRO with inputs of sample means
and sample covariance matrices. We name this method QUADRO-0 to
differentiate it from QUADRO.

%
%
\begin{figure}

\includegraphics{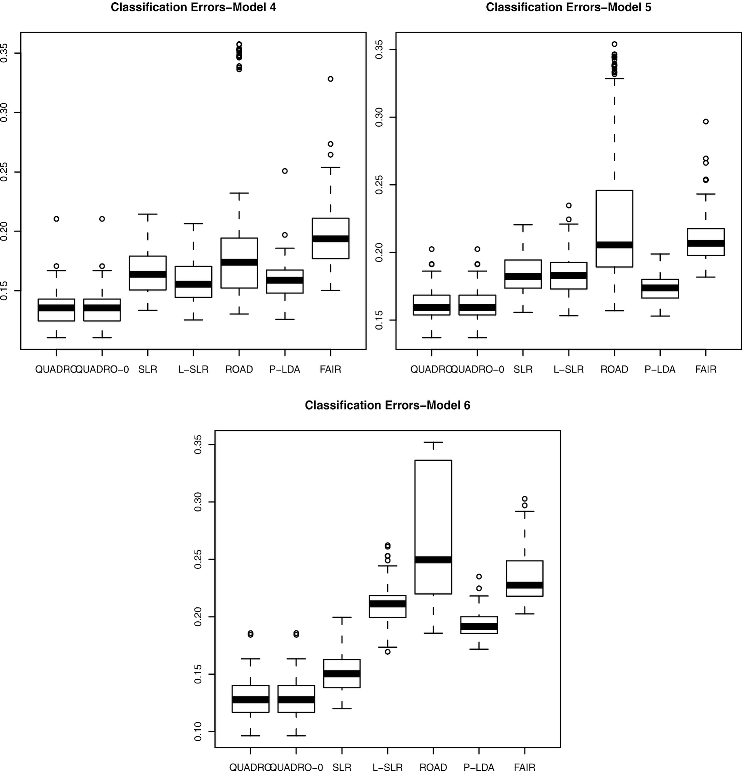}

\caption{Distributions of minimum classification error based on 100
replications across different elliptical distribution models. The
tuning parameters for QUADRO, SLR and L-SLR are chosen to minimize the
classification errors. See \citet{QUADROsupp} for detailed numerical tables.}
\label{figClassiErrorEllip}
\end{figure}

We consider three models:
\begin{itemize}[--]
\item[--] \textit{Model} 4: Here we use same parameters as those in {model~1}.
\item[--] \textit{Model} 5: $\bSigma_1$, $\bmu_1$ and $\bmu_2$ are the
same as in {model~1}. $\bSigma_2$ is the covariance matrix of a
fractional white noise process, where the difference parameter $l=0.2$.
In other words, $\bSigma_2$ has the polynomial off-diagonal decay
$\llvert \Sigma_2(i,j)\rrvert = O(\llvert i-j\rrvert ^{1-2l})$.
\item[--] \textit{Model} 6: $\bSigma_1$, $\bmu_1$ and $\bmu_2$ are the
same as in {model} 1. $\bSigma_2$ is a matrix such that $\Sigma
_2(i,j)=0.6^{\llvert i-j\rrvert }$; that is, $\bSigma_2$ has an exponential
off-diagonal decay.
\end{itemize}

Figure~\ref{figClassiErrorEllip} contains the boxplots of average
classification error over $100$ replications. QUADRO outperforms the
other methods in all settings. Also, QUADRO is better than QUADRO-0
(e.g., $0.161$ versus $0.173$, of the average classification error in
model~5), which illustrates the advantage of using the robust
estimators for means and covariance matrices.

\subsection{Real data analysis}

We apply QUADRO to a large-scale genomic dataset, GPL96, and compare
the performance of QUADRO
with SLR, L-SLR, ROAD, P-LDA and FAIR. The GPL96 data set contains
20,263 probes and 8124 samples from 309 tissues. Among the tissues,
breast tumor has 1142 samples, which is the largest set. We merge the
probes from the same gene by averaging them, and finally get 12,679
genes and 8124 samples.
We divide all samples into two groups: breast tumor or nonbreast tumor.

First, we look at the classification errors.
We replicate our experiment 100 times. Each time,
we proceed with the following steps:
\begin{itemize}
\item Randomly choose a training set of 400 samples, 200 from breast
tumor and 200 from nonbreast tumor.
\item For each training set, we use half of the samples to compute
$(\widehat{\bOmega},\widehat{\bdelta})$ and the other half to
select the tuning parameters by minimizing the classification error.
%
\item Use the remaining 942 samples from breast tumor and another
randomly chosen 942 samples from nonbreast tumor as testing set, and
calculate the testing error.
\end{itemize}
FAIR does not have any tuning parameters, so we use the whole training
set to calculate classification frontier, and the rest to calculate
testing error. The results are summarized in Table~\ref
{tbreal-classerr}. We see that QUADRO outperforms all other methods.

%
%
\begin{table}
\tabcolsep=0pt
\caption{Classification errors on GPL96 dataset, across methods
QUADRO, SLR and L-SLR. Means and standard deviations (in the
parenthesis) of $100$ replications are reported} \label{tbreal-classerr}
\begin{tabular*}{\tablewidth}{@{\extracolsep{\fill}}@{}lccccc@{}}
\hline
\textbf{QUADRO} & \textbf{SLR} & \textbf{L-SLR} & \textbf{ROAD} & \textbf{Penalized-LDA} & \textbf{FAIR}\\
\hline
0.014& 0.025 & 0.025& 0.016&0.060&0.046\\
(0.007)& (0.007)&(0.009)&(0.007)&(0.011)&(0.009)\\
\hline
\end{tabular*}
\end{table}

\begin{figure}[b]
\begin{tabular}{@{}cc@{}}

\includegraphics{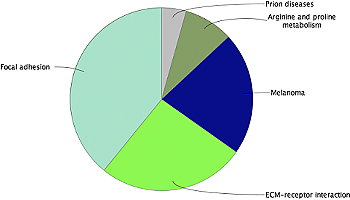}
 & \includegraphics{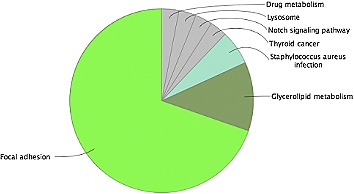}\\
\footnotesize{(a) QUADRO pathways} & \footnotesize{(b) SLR pathways}
\end{tabular}
\caption{Overall KEGG enrichment chart, using \textup{(a)} QUADRO; \textup{(b)} SLR.}
\label{figEnrichment}
\end{figure}

Next, we look at gene selection and focus on the two quadratic methods,
QUADRO and SLR. We apply two-fold cross-validation to both QUADRO and
SLR. In the results, QUADRO selects 139 genes and SLR selects 128 genes.
According to KEGG database, genes selected by QUADRO belong to 5 of the
pathways that contain more than two genes; correspondingly, genes
selected by SLR belong to 7 pathways. Using the ClueGo tool
[\citet{bindea2009cluego}], we display the overall KEGG
enrichment chart in
Figure~\ref{figEnrichment}. We see from Figure~\ref{figEnrichment} that both QUADRO and SLR have \textit{focal adhesion} as its
most important functional group. Nevertheless, QUADRO finds \textit{ECM-receptor interaction} as another important functional group.
\textit{ECM-receptor interaction} is a class consisting of a mixture of
structural and functional macromolecules, and it plays an important
role in maintaining cell and tissue structures and functions. Massive
studies [\citet{luparello2013aspects,wei2007markov}] have found
evidence that this class is closely related to breast cancer.

%
%

Besides the pathway analysis, we also perform the Gene Ontology (GO)
enrichment analysis on genes selected by QUADRO. This analysis was
completed by DAVID Bioinformatics Resources, and the results are shown
in Table~\ref{GO-Quadro}. We present the biological processes with
$p$-values smaller than $10^{-3}$. According to the table, we see that
many biological processes are significantly enriched, and they are
related to previously selected pathways. For instance, the biological
process \textit{cell adhesion} is known to be highly related to \textit{cell
communication pathways}, including \textit{focal adhesion} and \textit{ECM-receptor interaction}.

%
%
\begin{table}
\tabcolsep=0pt
\caption{Enrichment analysis results according to Gene Ontology for
genes selected by QUADRO. The four columns represent GO ID, GO
attribute, number of selected genes having the attribute and their
corresponding $p$-values. We rank them according to $p$-values in
increasing order} \label{GO-Quadro}
\begin{tabular*}{\tablewidth}{@{\extracolsep{\fill}}@{}lcd{2.0}d{1.6}@{\hspace*{-3.5pt}}}
\hline
\textbf{GO ID} & \textbf{GO attribute} & \multicolumn{1}{c}{\textbf{No. of genes}} & \multicolumn{1}{c}{\textbf{$\bolds{p}$-value}} \\
\hline
0048856& Anatomical structure development & 58 & 3.7\mbox{E--}12 \\
0032502&Developmental process&62&2.9\mbox{E--}10\\
0048731&System development&52&3.1\mbox{E--}10\\
0007275&Multicellular organismal development&55&1.8\mbox{E--}8\\
0001501&Skeletal system development&15&1.3\mbox{E--}6\\
0032501&Multicellular organismal process&66&1.4\mbox{E--}6\\
0048513&Organ development&37&1.4\mbox{E--}6\\
0009653&Anatomical structure morphogenesis&28&8.7\mbox{E--}6\\
0048869&Cellular developmental process&34&1.9\mbox{E--}5\\
0030154&Cell differentiation&33&2.1\mbox{E--}5\\
0007155&Cell adhesion&18 &2.4\mbox{E--}4\\
0022610&Biological adhesion&18 &2.2\mbox{E--}4\\
0042127&Regulation of cell proliferation &19 &2.9\mbox{E--}4\\
0009888&Tissue development&17&3.7\mbox{E--}4\\
0007398&Ectoderm development&9&4.8\mbox{E--}4\\
0048518&Positive regulation of biological process &34&5.6\mbox{E--}4\\
0009605&Response to external stimulus &20&6.3\mbox{E--}4\\
0043062&Extracellular structure organization &8&7.4\mbox{E--}4\\
0007399&Nervous system development&22&8.4\mbox{E--}4\\
\hline
\end{tabular*}
\end{table}

\section{Conclusions and extensions} \label{secconclude}

QUADRO is a robust sparse high-\break dimensional classifier, which allows us
to use differences in covariance matrices to enhance discriminability.
It is based on Rayleigh quotient optimization. The variance of
quadratic statistics involves all fourth cross moments, and this can
create both computational and statistical problems. These problems are
avoided by limiting our applications to the elliptical class of
distributions. Robust $M$-estimator and rank-based estimation of
correlations allow us to obtain the uniform convergence for
nonpolynomially many parameters, even when the underlying distributions
have the finite fourth moments. This allows us to establish oracle
inequalities under relatively weaker conditions.

Existing methods in the literature about constructing high-dimensional
quadra\-tic classifiers can be divided into two types. One is the
regularized QDA, where regularized estimates of $\bSigma^{-1}_1$ and
$\bSigma^{-1}_2$ are plugged into the Bayes classifier; see, for
example, \citet{friedman1989regularized}. QUADRO avoids directly
estimating inverse covariance matrices, which requires strong
assumptions in high dimensions. The other is to combine linear
classifiers with the inner-product kernel. The main difference between
QUADRO and this approach is the simplification in Proposition~\ref
{propellipmeanvar}. Due to this simplification, QUADRO avoids
incorporating all fourth cross moments from the data and gains extra
statistical efficiency.

QUADRO also has deep connections with the literature of sufficient
dimension reduction. Dimension reduction methods, such as SIR
[\citet{li1991sliced}], SAVE [\citet{cook1991comment}] and
Directional
Regression [\citet{li2007directional}], can be equivalently formulated
as maximizing some ``quotients.''
The population objective of SIR is to maximize $\var\{\mathbb
{E}[f(\Xb\mid Y)]\}$ subject to $\var[f(\Xb)]=1$.
Using the same
constraint, SAVE and directional regression combine $\var\{\mathbb
{E}[f(\Xb\mid Y)]\}$ and $\mathbb{E}[\var(f(\Xb\mid Y))]$ in the objective.
An interesting observation is that the Rayleigh quotient maximization
is equivalent to the population objective of SIR, by noting that the
denominator of (\ref{Rq}) is equal to $\mathbb{E}[\var(f(\Xb\mid Y))]$
and $\var[f(\Xb)]=\mathbb{E}[\var(f(\Xb\mid Y))]+ \var\{\mathbb
{E}[f(\Xb\mid Y)]\}$. This is not a coincidence, but due instead to the
known equivalence between SIR and LDA in classification [\citet
{kent1991discussion,Li00highdimensional}].

Despite similar population objectives, QUADRO and the aforementioned
dimension reduction methods are different in important ways. First, we
clarify that even when $\lambda_1,\lambda_2$ are $0$, QUADRO is not
the same procedure as SIR combined with the inner-product kernel
[\citet{wu2008kernel}], although they share the same population
objective. The
difference is that
QUADRO utilizes a simplification of the Rayleigh quotient for quadratic
$f$, relying on the assumption that $\Xb\mid Y$ is always elliptically
distributed; moreover, it adopts robust estimators of the mean vectors
and covariance matrices. Second, QUADRO is designed for
high-dimensional settings, in which neither SIR, SAVE nor Directional
Regression can be directly implemented. These methods need to either
standardize the original data $\Xb\mapsto\hbSigma^{-1}(\Xb-\bar
{\Xb})$ or solve a generalized eigen-decomposition problem $\bA\bv
=\lambda\hbSigma\bv$ for some matrix~$\bA$. Both methods require
that the sample covariance matrix is well conditioned, which is often
not the case in high dimensions. Possible solutions include Regularized
SIR [\citet{zhong2005rsir,li2008sliced}], solving generalized
eigen-decomposition for an undetermined system [\citet
{coudret2014comparison}] and variable selection approaches [\citet
{chen2010coordinate,jiang2013sliced}]. However, these methods are not
designed for Rayleigh quotient maximization. Third, our assumption on
the model is different from that in dimension reduction. We require
$\Xb\mid Y$ to be elliptically distributed, while many dimension reduction
methods ``implicitly'' require $\Xb$ to be marginally elliptically
distributed. Neither method is stronger than the other. Assuming
conditional elliptical distribution is more natural in classification.
In addition, our assumption is used only to simplify the variances of
quadratic statistics, whereas the elliptical assumption is critical to SIR.



The Rayleigh optimization framework developed in this paper can be
extended to the multi-class case. Suppose the data are drawn
independently from a joint distribution of $(\Xb, Y)$, where $\Xb\in
\mathbb{R}^d$ and $Y$ takes values in $\{0,1,\ldots, K-1\}$.
Definition (\ref{Rq}) for the Rayleigh quotient of a projection
$f\dvtx \mathbb{R}^d\to\mathbb{R}$ is still well defined. Let $\pi
_k=\mathbb{P}(Y=k)$, for $k=0, 1,\ldots, K-1$. In this $K$-class situation,
%
%
\begin{equation}
\label{Rq-multiclass} \Rq(f) = \frac{\sum_{0\leq k<l\leq K-1} \pi_k\pi
_l \{\mathbb
{E}[f(\Xb)\mid Y=k] - \mathbb{E}[f(\Xb)\mid Y=l] \}^2}{ \sum_{0\leq
k\leq K-1} \pi_k \var[f(\Xb)\mid Y=k]}.
\end{equation}
Let $M_k(f)=\mathbb{E}[f(\Xb)\mid Y=k]$ and $L_k(f)=\var[f(\Xb)\mid Y=k]$.
Similar to the two-class case, maximizing $\Rq(f)$ is equivalent to
solving the following optimization problem:
\[
\min_f \sum_{k=0}^{K-1}
\pi_k L_k(f)\quad\mbox{s.t.}\quad \sum
_{0\leq k<l\leq K-1}\pi_k\pi_l \bigl\llvert
M_k(f)-M_l(f) \bigr\rrvert^2= 1.
\]
However, this is not a convex problem. We consider an approximate
Rayleigh-quotient-maximization problem as follows:
\[
\min_f \sum_{k=0}^{K-1}
\pi_k L_k(f) \quad\mbox{s.t.}\quad\sqrt{
\pi_k\pi_l} \bigl\llvert M_k(f)-M_l(f)
\bigr\rrvert\geq1, \qquad0\leq k<l\leq K-1.
\]
To solve this problem, we first pick an order of $M_1(f),\ldots,
M_{K}(f)$ to remove the absolute values in the constraints. Then it
becomes a convex problem. Therefore, the whole optimization can be
carried out by simultaneously solving $K!$ convex problems. When $K$ is
small, the computational cost is reasonable. In practice, we can apply
more efficient algorithms to speed up the computation.

\section{Proofs} \label{secproof}

\subsection{Proof of Theorem \texorpdfstring{\protect\ref{teoRbound}}{5.1}}

We prove\vspace*{2pt} the claim by first rewriting optimization problem (\ref
{progquadro}) into a vector form. For any $(\bOmega,\bdelta)$, write
$\bx=[\vec(\bOmega)^{\top},\break  \bdelta^{\top}]^{\top}$. Let $\bQ$
be as defined in Section~\ref{secanalysis}, and
\[
\mathbf{q}= %
\lleft[ \matrix{ \vec\bigl(\bSigma_2+
\bmu_2\bmu_2^{\top} - \bSigma_1 -
\bmu_1\bmu_1^{\top} \bigr)^{\top},  2(
\bmu_1-\bmu_2)^{\top} } \rright]
^{\top}.
\]
We introduce the following lemma which is proved in the supplementary material [\citet{QUADROsupp}].
%

\begin{lem} \label{lemvectorize}
$M(\bOmega, \bdelta)=\mathbf{q}^{\top}\bx$ and $L(\bOmega,\bdelta
)=\bx^{\top}\bQ\bx$.
\end{lem}
%

Let $\bx^*_{\lambda_0}=[\vec(\bOmega^*_{\lambda
_0})^{\top}, (\bdelta^*_{\lambda_0})^{\top}]^{\top}$ and $\hbx
=[\vec(\hbOmega)^{\top}, \hbdelta^{\top}]^{\top}$. Using Lemma
\ref{lemvectorize},
\begin{eqnarray*}
\bx_{\lambda_0}^* &=& \min_{\bx\in\mathbb{R}^d\dvtx  \mathbf{q}^{\top}\bx
=1} \bigl\{\bx^{\top}
\bQ\bx+ \lambda_0 \llvert\bx\rrvert_1 \bigr\},
\\
\hbx& = &\argmin_{\bx\in\mathbb{R}^d\dvtx  \widehat{\mathbf{q}}^{\top}\bx
=1} \bigl\{ \bx^{\top}\hbQ\bx+ \lambda\llvert
\bx\rrvert_1 \bigr\},
\end{eqnarray*}
where $\hbQ$ and $\widehat{\mathbf{q}}$ are counterparts of $\bQ$ and
$\mathbf{q}$, respectively, by replacing $\bmu_1$, $\bmu_2$, $\bSigma_1$
and $\bSigma_2$ with their estimates. Moreover, we have the Rayleigh quotient
\[
R(\bOmega, \bdelta)=R(\bx)\equiv\frac{(\mathbf{q}^{\top}\bx
)^{2}}{\bx^{\top}\bQ\bx}.
\]
In addition, we have the following lemma, which is proved in
the supplementary material [\citet{QUADROsupp}].
%

\begin{lem} \label{leminfnorm}
$\max\{\llvert \hbQ-\bQ\rrvert _\infty, \llvert \widehat{\mathbf
{q}}-\mathbf{q}\rrvert _\infty\}
\leq C_0 \max\{\llvert \hbSigma_k-\bSigma_k\rrvert _\infty, \llvert
\hbmu_k-\bmu
_k\rrvert _\infty, k=1,2\}$ for some constant $C_0>0$.\vadjust{\goodbreak}
\end{lem}
%

Combining the above results, the claim follows immediately from the
following theorem:
%

\begin{teo} \label{teotechnique}
For any $\lambda_0\geq0$, let $S$ be the support of $\bx^*_{\lambda
_0}$. Suppose $\Theta(S, 0)\geq c_0$, $\Theta(S, 3)\geq a_0$ and
$R(\bx^*_{\lambda_0})\geq u_0$, for positive constants $a_0$, $c_0$
and~$u_0$. Let $\Delta_n=\max\{\llvert \hbQ-\bQ\rrvert _\infty,
\llvert \widehat{\mathbf{q}}- \mathbf{q}\rrvert _\infty\}$,
$s_0=\llvert S\rrvert $ and $k_0=\max\{s_0,\break  R(\bx
^*_{\lambda_0})\}$. Suppose $4s_0\Delta_n^2<c_0u_0$ and $\max\{
s_0\Delta_n, s_0^{1/2}k_0^{1/2}\lambda_0 \}<1$. Then there exist
positive constants $C=C(a_0, c_0, u_0)$ and $A= A(a_0, c_0, u_0)$, such\vspace*{1pt}
that for any $\eta>1$, by taking $\lambda=C\eta\max\{
s_0^{1/2}\Delta_n, k_0^{1/2}\lambda_0 \}{[R(\bx^*_{\lambda_0})]^{-1/2}}$,
\[
\frac{R(\hbx)}{R(\bx^*_{\lambda_0})} \geq1 - A\eta^2 \max\bigl\{ s_0
\Delta_n, s_0^{1/2}k_0^{1/2}
\lambda_0 \bigr\}.
\]
\end{teo}
%

The main part of the proof is to show Theorem \ref{teotechnique}.
Write for short $\bx^*=\bx^*_{\lambda_0}$, $R^*=R(\bx^*)$,
$V^*=(R^*)^{-1}= (\bx^*)^{\top}\bQ\bx^*$, $\bar{V}^*=(V^*)^{1/2}$.
Let $\alpha_n = \Delta_n \llvert \bx^*\rrvert _0^{1/2}$, $\beta_n =
\Delta_n
\llvert \bx^*\rrvert _0$ and $T_n(\bx^*) = \max\{s_0 \Delta_n,
s_0^{1/2}k_0^{1/2}\lambda_0\}$. {We define} the quantity
\[
\Gamma(\bx) = \frac{\llvert \bQ\bx- (\bx^{\top}\bQ\bx) \mathbf
{q}\rrvert _\infty}{(\bx^{\top}\bQ\bx)^{1/2}}\qquad\mbox{for any }\bx.
\]

\begin{longlist}[\textit{Step} 2.]
\item[\textit{Step} 1.] We introduce $\bx^*_1$, a multiple of $\bx^*$, and use
it to bound $\llvert \hbx\rrvert _1$.

Let $\bQ_{SS}$ be the submatrix of $\bQ$ formed by rows and columns
corresponding to~$S$. Since $\lambda_{\min}(\bQ_{SS}){=\Theta
(S,0)}\geq c_0$, we have $(\bx^*)^{\top}\bQ\bx^*\geq c_0\llvert \bx
^*\rrvert ^2$. Using this fact and by the Cauchy--Schwarz inequality,
%
%
\begin{equation}
\label{qQtemp0} \bigl\llvert\bx^* \bigr\rrvert_1 \leq\sqrt{ \bigl
\llvert\bx^* \bigr\rrvert_0} \bigl\llvert\bx^* \bigr\rrvert\leq
c_0^{-1/2} \sqrt{ \bigl\llvert\bx^* \bigr\rrvert
_0} \bar{V}^*.
\end{equation}
It follows that
%
%
\begin{equation}
\label{qQtemp1} \bigl\llvert\widehat{\mathbf{q}}^{\top}\bx^*-
\mathbf{q}^{\top}\bx^* \bigr\rrvert\leq\llvert\widehat{\mathbf{q}}-
\mathbf{q}\rrvert_{\infty} \bigl\llvert\bx^* \bigr\rrvert_1
\leq c_0^{-1/2} \Delta_n \sqrt{ \bigl\llvert
\bx^* \bigr\rrvert_0} \bar{V}^* = c_0^{-1/2}
\alpha_n\bar{V}^*.
\end{equation}
{Let}\vspace*{1pt} $t_n=\widehat{\mathbf{q}}^{\top}\bx^*$. Then (\ref{qQtemp1})
says that $\llvert t_n-1\rrvert \leq c_0^{-1/2}\alpha_n \bar{V}^*$.
{Noting that
$\bar{V}^*=(R^*)^{1/2}\leq u_0^{-1/2}$, we have $\llvert t_n-1\rrvert
\leq
(c_0u_0)^{-1/2}s_0^{1/2}\Delta_n<1/2$ by assumption.} In particular,
$t_n>0$. Let
\[
\bx^*_1=t_n^{-1}\bx^*.
\]
Then $\widehat{\mathbf{q}}^{\top}\bx^*_1=1$. From the definition of
$\hbx$,
%
%
\begin{equation}
\label{qQtemp2} \hbx^{\top}\hbQ\hbx+ \lambda\llvert\hbx\rrvert
_1 \leq\bigl(\bx^*_1 \bigr)^{\top}\hbQ
\bx^*_1 + \lambda\bigl\llvert\bx^*_1 \bigr\rrvert
_1.
\end{equation}
By direct calculation,
%
%
\begin{eqnarray}
\label{qQtemp31} \hbx^{\top}\hbQ\hbx- \bigl(\bx^*_1
\bigr)^{\top}\hbQ\bx^*_1 &=& \bigl(\hbx-\bx^*_1
\bigr)^{\top}\hbQ\bigl(\hbx-\bx^*_1 \bigr) + 2 \bigl(\hbx-
\bx^*_1 \bigr)^{\top}\hbQ\bx^*_1\nonumber
\\
&=& \bigl(\hbx-\bx^*_1 \bigr)^{\top}\hbQ\bigl(\hbx-
\bx^*_1 \bigr) + 2 \bigl(\hbx-\bx^*_1
\bigr)^{\top} \bigl(\hbQ\bx^*_1- V^* \widehat{\mathbf{q}}
\bigr)
\\
& \geq& 2 \bigl(\hbx-\bx^*_1 \bigr)^{\top} \bigl(
\hbQ\bx^*_1- V^* \widehat{\mathbf{q}} \bigr),
\nonumber
\end{eqnarray}
where the second equality is due to $\widehat{\mathbf{q}}^{\top}\hbx
=\widehat{\mathbf{q}}^{\top}\bx^*_1=1$. {We aim to bound $\llvert \hbQ
\bx
^*_1-{V^*}\widehat{\mathbf{q}}\rrvert _\infty$. The following lemma is proved
in the supplementary material [\citet{QUADROsupp}].}
%

\begin{lem} \label{lemKKTsmall}
When $\Theta(S, 0)\geq c_0$, there exists a positive constant
$C_1=C_1(c_0)$ such that $\Gamma(\bx^*_{\lambda_0})\leq C_1 \lambda
_0 [\max\{ s_0, R(\bx^*_{\lambda_0}) \}]^{1/2}$ for any $\lambda
_0\geq0$.
\end{lem}
%

Since $\bx^*_1=t_n^{-1}\bx^*$ and $t_n^{-1}< 2$,
\begin{eqnarray*}
&& \bigl\llvert\hbQ\bx^*_1-{V^*}\widehat{\mathbf{q}} \bigr\rrvert
_\infty
\\
&&\qquad \leq t_n^{-1} \bigl\llvert\hbQ\bx^*- V^*
\widehat{\mathbf{q}} \bigr\rrvert_\infty+ V^* \bigl\llvert
t_n^{-1}-1 \bigr\rrvert\llvert\widehat{\mathbf{q}}\rrvert
_\infty
\\
&&\qquad \leq 2 \bigl( \bigl\llvert\bQ\bx^*- V^*\mathbf{q} \bigr
\rrvert_\infty+ \llvert\hbQ-\bQ\rrvert_\infty\bigl\llvert
\bx^* \bigr\rrvert_1 + V^* \llvert\widehat{\mathbf{q}}-\mathbf{q}
\rrvert_\infty+ V^* \llvert t_n-1\rrvert\llvert\widehat{
\mathbf{q}}\rrvert_\infty\bigr)
\\
&&\qquad \leq 2 \bigl[ \Gamma\bigl(\bx^*
\bigr)\bar{V}^* + c_0^{-1/2}\alpha_n\bar{V}^* +
{u_0^{-1/2} \Delta_n\bar{V}^*} + {\llvert
\widehat{\mathbf{q}}\rrvert_\infty}c_0^{-1/2}u_0^{-1}
\alpha_n\bar{V}^* \bigr]
\\
&&\qquad \leq  C_2 \bigl(
\lambda_0k_0^{1/2} + s_0^{1/2}
\Delta_n \bigr){\bar{V}^*}.
\end{eqnarray*}
Here the third inequality follows from (\ref{qQtemp0})--(\ref
{qQtemp1}) and {$V^*=\bar{V}^*(R^*)^{-1/2}\leq u_0^{-1/2}\bar{V}^*$}.
The last inequality is obtained as follows: from Lemma \ref
{leminfnorm}, we know that $\llvert \widehat{\mathbf{q}}\rrvert
_\infty\leq\llvert \mathbf{q}\rrvert _\infty+\llvert \widehat{\mathbf
{q}}-\mathbf{q}\rrvert _\infty\leq2C_0$ (see also
the assumptions in the beginning of Section~\ref{subsecoracle}); we
also use Lemma \ref{lemKKTsmall} and $\alpha_n\bar{V}^*\leq
u_0^{-1/2} s_0^{1/2}\Delta_n$.
By letting $C=8C_2$, the choice of $\lambda=C\eta\max\{
s_0^{1/2}\Delta_n, k_0^{1/2}\lambda_0\}{\bar{V}^*}$ for $\eta>1$
ensures that
\[
\bigl\llvert\hbQ\bx^*_1-\widehat{\mathbf{q}} \bigr\rrvert
_\infty\leq\lambda/4.
\]
Plugging this result into (\ref{qQtemp31}) gives
%
%
\begin{equation}
\label{qQtemp3} \hbx^{\top}\hbQ\hbx- \bigl(\bx^*_1
\bigr)^{\top}\hbQ\bx^*_1\geq- \frac
{\lambda}{2} \bigl\llvert
\hbx-\bx^*_1 \bigr\rrvert_1.
\end{equation}
Combining (\ref{qQtemp2}) and (\ref{qQtemp3}) gives
%
%
\begin{equation}
\label{qQtemp4origin} \lambda\llvert\hbx\rrvert_1 - \frac{\lambda}{2}
\bigl\llvert\hbx- \bx_1^* \bigr\rrvert_1 \leq\lambda
\bigl\llvert\bx_1^* \bigr\rrvert_1.
\end{equation}
First, since $\llvert \hbx\rrvert _1=\llvert \hbx_S\rrvert _1 +
\llvert \hbx_{S^c}\rrvert _1\geq\llvert \bx
_{1S}^*\rrvert _1 - \llvert \hbx_S-\bx^*_{1S}\rrvert _1 + \llvert
\hbx_{S^c}\rrvert _1$ and $\llvert \hbx-\bx
_1^*\rrvert _1=\llvert \hbx_{S}-\bx^*_{1S}\rrvert _1 + \llvert \hbx
_{S^c}\rrvert _1$, we immediately
see from (\ref{qQtemp4origin}) that
%
%
\begin{equation}
\label{qQtempRE} \bigl\llvert\bigl(\hbx-\bx^*_1
\bigr)_{S^c} \bigr\rrvert_1 \leq3 \bigl\llvert\bigl(\hbx-
\bx^*_1 \bigr)_S \bigr\rrvert_1.
\end{equation}
Second, note that $\llvert \hbx-\bx_1^*\rrvert _1\leq\llvert \hbx
\rrvert _1 + \llvert \bx_1^*\rrvert _1$.
Plugging this into (\ref{qQtemp4origin}) gives
%
%
\begin{equation}
\label{qQtemp4} \llvert\hbx\rrvert_1 \leq3 \bigl\llvert
\bx^*_1 \bigr\rrvert_1 = 3t_n^{-1}
\bigl\llvert\bx^* \bigr\rrvert_1 \leq6c_0^{-1/2}
\sqrt{ \bigl\llvert\bx^* \bigr\rrvert_0} \bar{V}^*.
\end{equation}
\end{longlist}

\begin{longlist}[\textit{Step} 2.]
\item[\textit{Step} 2.] We use (\ref{qQtempRE})--(\ref{qQtemp4}) to
derive an upper bound for $(\hbx)^{\top}\bQ\hbx- (\bx_1^*)^{\top
}\bQ\bx_1^*$.

Note that
%
%
\begin{eqnarray}
\label{qQtemp5}
\quad && \hbx^{\top}\hbQ\hbx- \bigl(\bx^*_1
\bigr)^{\top}\hbQ\bx^*_1\nonumber
\\
&&\qquad \geq \hbx^{\top}\bQ\hbx- \bigl(\bx^*_1
\bigr)^{\top}\bQ\bx^*_1 - \bigl( \bigl\llvert
\hbx^{\top}\hbQ\hbx- \hbx^{\top}\bQ\hbx\bigr\rrvert+ \bigl
\llvert\bigl(\bx^*_1 \bigr)^{\top
}\hbQ\bx^*_1 -
\bigl( \bx_1^* \bigr)^{\top}\bQ\bx^*_1 \bigr
\rrvert\bigr)
\nonumber
\\
&&\qquad \geq \hbx^{\top}\bQ\hbx- \bigl(\bx^*_1
\bigr)^{\top}\bQ\bx^*_1 - \bigl(\llvert\hbQ-\bQ\rrvert
_\infty\llvert\hbx\rrvert_1^2 + \llvert\hbQ-
\bQ\rrvert_\infty\bigl\llvert\bx^*_1 \bigr\rrvert
_1^2 \bigr)
\\
&&\qquad \geq \hbx^{\top}\bQ\hbx- \bigl(\bx^*_1
\bigr)^{\top}\bQ\bx^*_1 - 10t_n^{-{2}}
\llvert\hbQ-\bQ\rrvert_\infty\bigl\llvert\bx^* \bigr\rrvert
_1^2
\nonumber
\\
&&\qquad \geq \hbx^{\top}\bQ\hbx- \bigl(\bx^*_1
\bigr)^{\top}\bQ\bx^*_1 - C_3 \beta_n
V^*,
\nonumber
\end{eqnarray}
where the last two inequalities are direct results of (\ref{qQtemp4}).
Combining (\ref{qQtemp2}) and (\ref{qQtemp5}),
%
%
\begin{equation}
\label{qQtemp6} \hbx^{\top}\bQ\hbx+ \lambda\llvert\hbx\rrvert
_1 \leq\bigl(\bx_1^* \bigr)^{\top}\bQ
\bx_1^* + \lambda\bigl\llvert\bx_1^* \bigr\rrvert
_1 + C_3 \beta_n V^*.
\end{equation}
Similar to (\ref{qQtemp31}), we have
%
%
\begin{equation}
\label{qQtemp11}
\quad \hbx^{\top}\bQ\hbx- \bigl(\bx^*_1
\bigr)^{\top}\bQ\bx^*_1 = \bigl(\hbx-\bx^*_1
\bigr)^{\top}\bQ\bigl(\hbx-\bx^*_1 \bigr) + 2 \bigl(\hbx-
\bx^*_1 \bigr)^{\top} \bigl(\bQ\bx^*_1- V^*
\widehat{\mathbf{q}} \bigr),
\end{equation}
where
\begin{eqnarray*}
\bigl\llvert\bQ\bx_1^*-V^*\widehat{\mathbf{q}} \bigr\rrvert
_\infty&\leq& t_n^{-1} \bigl( \bigl\llvert\bQ
\bx^*- V^*\mathbf{q} \bigr\rrvert_\infty+ V^*\llvert\widehat{
\mathbf{q}}-\mathbf{q}\rrvert_\infty\bigr) + V^* \bigl\llvert
t_n^{-1}-1 \bigr\rrvert\llvert\widehat{\mathbf{q}}\rrvert
_\infty
\\
&\leq&2 \bigl[ \Gamma\bigl(\bx^* \bigr){\bar{V}^*} +
{u_0^{-1/2}\Delta_n\bar{V}^*} + {\llvert
\widehat{\mathbf{q}}\rrvert_\infty c_0^{-1/2}u_0^{-1}
\alpha_n\bar{V}^*} \bigr]
\\
&\leq&\lambda/4.
\end{eqnarray*}
It follows that
\[
\hbx^{\top}\bQ\hbx- \bigl(\bx^*_1 \bigr)^{\top}\bQ
\bx^*_1 \geq\bigl(\hbx-\bx^*_1 \bigr)^{\top}\bQ
\bigl(\hbx-\bx^*_1 \bigr) - \frac{\lambda
}{2} \bigl\llvert\hbx-
\bx^*_1 \bigr\rrvert_1.
\]
Plugging this into (\ref{qQtemp6}), we obtain
%
%
\begin{equation}
\label{qQtemp7} \bigl(\hbx-\bx^*_1 \bigr)^{\top}\bQ\bigl(
\hbx- \bx^*_1 \bigr) + \lambda\llvert\hbx\rrvert_1 -
\frac{\lambda}{2} \bigl\llvert\hbx-\bx^*_1 \bigr\rrvert
_1 \leq\lambda\bigl\llvert\bx^*_1 \bigr\rrvert
_1 + C_3 \beta_n V^*.
\end{equation}
We can rewrite the second and third terms on the left-hand side of
(\ref{qQtemp7}) as
\[
\lambda\llvert\hbx_S\rrvert_1 - \frac{\lambda}{2}
\bigl\llvert\hbx_S - \bx^*_{1S} \bigr\rrvert
_1 + \frac{\lambda}{2}\llvert\hbx_{S^c}\rrvert
_1.
\]
Plugging this into (\ref{qQtemp7}) and by the triangular inequality
$\llvert \bx_{1S}^*\rrvert _1 - \llvert \hbx_S\rrvert _1\leq
\llvert \hbx_S-\bx^*_{1S}\rrvert _1$, we find that
\[
\bigl(\hbx-\bx^*_1 \bigr)^{\top}\bQ\bigl(\hbx-
\bx^*_1 \bigr) + \frac{\lambda}{2}\llvert\hbx_{S^c}
\rrvert_1 \leq\frac{3\lambda}{2} \bigl\llvert\hbx_{S}-
\bx^*_{1S} \bigr\rrvert_1 + C_3
\beta_n V^*.
\]
We drop the term $\frac{\lambda}{2}\llvert \hbx_{S^c}\rrvert _1$ on
the left-hand
side and apply the Cauchy--Schwarz inequality to the term $\llvert
\hbx
_{S}-\bx^*_{1S}\rrvert _1$. This gives
%
%
\begin{eqnarray}
\label{qQtemp8} \bigl(\hbx-\bx^*_1 \bigr)^{\top}\bQ\bigl(
\hbx-\bx^*_1 \bigr) &\leq&\frac{3\lambda
}{2} \sqrt{ \bigl\llvert
\bx^*_1 \bigr\rrvert_0} \bigl\llvert
\hbx_{1S}-\bx^*_{1S} \bigr\rrvert+ C_3
\beta_n V^*.
\end{eqnarray}
Since (\ref{qQtempRE}) holds, { by the definition of $\Theta(S,3)$, }
\[
\bigl(\hbx-\bx_1^* \bigr)^{\top}\bQ\bigl(\hbx-
\bx_1^* \bigr)\geq a_0 \bigl\llvert\hbx_S-
\bx^*_{1S} \bigr\rrvert^2.
\]
We write temporarily $Y=(\hbx-\bx_1^*)^{\top}\bQ(\hbx-\bx_1^*)$
and $b=C_3 \beta_n V^*$. Combining these with (\ref{qQtemp8}),
\[
Y \leq\frac{3\lambda}{2\sqrt{a_0}}\sqrt{ \bigl\llvert\bx^*_1 \bigr
\rrvert
_0 Y} + b.
\]
Note that when $u^2\leq au+b$, we have $(u-\frac{a}{2})^2\leq b +
\frac{a^2}{4}$, and hence $u^2\leq2[\frac{a^2}{4} + (u-\frac
{a}{2})^2]\leq a^2 + 2b$. As a result, the above inequality implies
%
%
\begin{equation}
\label{qQtemp121} \bigl(\hbx-\bx^*_1 \bigr)^{\top}\bQ\bigl(
\hbx- \bx^*_1 \bigr) \leq\frac{9\lambda
^2}{4a_0} \bigl\llvert\bx^* \bigr
\rrvert_0 + 2C_3 \beta_n V^*,
\end{equation}
{where we have used $\llvert \bx^*_1\rrvert _0=\llvert \bx^*\rrvert _0$.}
Furthermore, (\ref{qQtemp11}) yields that
%
%
\begin{eqnarray}
\label{qQtemp122} \hbx^{\top}\bQ\hbx- \bigl(\bx^*_1
\bigr)^{\top}\bQ\bx^*_1 &\leq&\bigl(\hbx-\bx^*_1
\bigr)^{\top}\bQ\bigl(\hbx-\bx^*_1 \bigr) +
\frac{\lambda
}{2} \bigl\llvert\hbx-\bx^*_1 \bigr\rrvert
_1\nonumber
\\
&\leq&\bigl( \hbx-\bx^*_1 \bigr)^{\top}
\bQ\bigl(\hbx-\bx^*_1 \bigr) + 2\lambda\bigl\llvert\bx
^*_1 \bigr\rrvert_1
\\
&\leq&\bigl(\hbx-
\bx^*_1 \bigr)^{\top}\bQ\bigl(\hbx- \bx^*_1
\bigr) + 4 c_0^{-1/2}\bar{V}^* \lambda\sqrt{ \bigl\llvert
\bx^* \bigr\rrvert_0},\nonumber
\end{eqnarray}
where\vspace*{1pt} the second inequality is due to $\llvert \hbx-\bx_1^*\rrvert
_1\leq\llvert \hbx
\rrvert _1+\llvert \bx^*_1\rrvert \leq4\llvert \bx_1^*\rrvert _1$,
and the last inequality is from
(\ref{qQtemp4}).
Recall that $\lambda=C\eta\max\{k_0^{1/2}\lambda_0, s_0^{1/2}\Delta
_n\} \bar{V}^*$. As a result,
%
%
\begin{equation}
\label{qQtemp123} \lambda\sqrt{ \bigl\llvert\bx^* \bigr\rrvert_0} =
C \eta\max\bigl\{ k_0^{1/2}s_0^{1/2}
\lambda_0, s_0 \Delta_n \bigr\} \bar{V}^*=
C \eta T_n \bigl(\bx^* \bigr) \bar{V}^*.
\end{equation}
Combining (\ref{qQtemp121}), (\ref{qQtemp122}) and (\ref
{qQtemp123}) gives
%
%
\begin{eqnarray}
\label{qQtemp13} && \hbx^{\top}\bQ\hbx- \bigl(\bx^*_1
\bigr)^{\top}\bQ\bx^*_1\nonumber
\\
&&\qquad \leq \frac{9C^2}{4a_0}
\eta^2 \bigl[T_n \bigl(\bx^* \bigr) \bigr]^2
V^* + 4C c_0^{-1/2}\eta T_n \bigl(\bx^* \bigr)
V^* + 2C_3 \beta_n V^*
\\
&&\qquad\leq C_4
\eta^2 T_n \bigl(\bx^* \bigr) V^*.\nonumber
\end{eqnarray}
\end{longlist}

\begin{longlist}[\textit{Step} 2.]
\item[\textit{Step} 3.] We use (\ref{qQtemp13}) to give a lower bound
of $R(\hbx)$.

Note that $R(\hbx)=(\mathbf{q}^{\top}\hbx)^2/(\hbx^{\top}\bQ\hbx)$.
First, we look at the denominator $\hbx^{\top}\bQ\hbx$. From (\ref
{qQtemp1}) and that $t_n>1/2$,
\[
\bigl\llvert t_n^{-2}-1 \bigr\rrvert=
t_n^{-1} \bigl(1+t_n^{-1} \bigr)
\llvert t_n-1\rrvert\leq6 c_0^{-1/2}\alpha
_n\bar{V}^*.
\]
Combining with (\ref{qQtemp13}) and noting that $(\bx_1^*)^{\top}\bQ
\bx_1^*=t_n^{-2}(\bx^*)^{\top}\bQ\bx^*=t_n^{-2}V^*$, we have
%
%
\begin{eqnarray}
\label{qQtemp16} \hbx^{\top}\bQ\hbx&\leq&\bigl[t_n^{-2}
+ C_4 \eta^2 T_n \bigl(\bx^* \bigr) \bigr]
\bigl(\bx^* \bigr)^{\top}\bQ\bx^*\nonumber
\\
&\leq&\bigl[1 + 6
c_0^{-1/2}\alpha_n\bar{V}^* + C_{4}
\eta^2 T_n \bigl(\bx^* \bigr) \bigr] \bigl(\bx^*
\bigr)^{\top}\bQ\bx^*
\\
&\leq&\bigl[1 + C_5
\eta^2 T_n \bigl(\bx^* \bigr) \bigr] \bigl(\bx^*
\bigr)^{\top}\bQ\bx^*.\nonumber
\end{eqnarray}
Second, we look at the numerator $\mathbf{q}^{\top}\hbx$. Since
$\widehat{\mathbf{q}}^{\top}\hbx=1$, {by (\ref{qQtemp4}),}
%
%
\begin{equation}
\label{qQtemp17} \bigl\llvert\mathbf{q}^{\top}\hbx- 1 \bigr\rrvert\leq
\llvert\widehat{\mathbf{q}}-\mathbf{q}\rrvert_\infty\llvert\hbx
\rrvert_1 \leq6c_0^{-1/2} \alpha_n
\bar{V}^* \leq C_6 T_n \bigl(\bx^* \bigr).
\end{equation}
Combining (\ref{qQtemp16}) and (\ref{qQtemp17}) gives
%
%
\begin{eqnarray}
R(\hbx) &=& \frac{(\mathbf{q}^{\top}\hbx)^2}{\hbx^{\top}\bQ\hbx} \geq
\frac{ [1- C_6T_n(\bx^*) ]^2}{1 + C_{5} \eta^2 T_n(\bx
^*)} \frac{1}{(\bx^*)^{\top} \bQ\bx^*}\nonumber
\\
&\geq&\bigl[1 - A\eta^2 T_n \bigl(\bx^* \bigr) \bigr]
\frac{(\mathbf{q}^{\top}\bx
^*)^2}{(\bx^*)^{\top}\bQ\bx^*}
\\
& =& \bigl[1 - A\eta^2 T_n \bigl(
\bx^* \bigr) \bigr] R \bigl(\bx^* \bigr),\nonumber
\end{eqnarray}
where $A=A(a_0, c_0, u_0)$ is a positive constant.
\end{longlist}

\subsection{Proof of Proposition \texorpdfstring{\protect\ref{properrgap}}{6.1}}

Denote by $\mathbb{P}(i\mid j)$ the probability that a new sample from
class $j$ is misclassified to class $i$, for $i,j\in\{1,2\}$ and
$i\neq j$. The classification error of $h$ is
\[
\operatorname{err}(h) = \pi\mathbb{P}(2\mid1)+ (1- \pi)\mathbb
{P}(1\mid2).
\]
Write $M_k=M_k(\bOmega,\bdelta)$ and $L_k=L_k(\bOmega,\bdelta)$ for
short. It suffices to show that
\begin{eqnarray*}
\mathbb{P}(2\mid1) &=&\bar{\Phi} \biggl( \frac{(1-t)M}{\sqrt
{L_1}} \biggr) +
\frac{O(q) + o(d)}{L_1^{3/2}},
\\
\mathbb{P}(1\mid2) &=&\bar{\Phi} \biggl(
\frac{tM}{\sqrt{L_2}} \biggr) + \frac{O(q) + o(d)}{L_2^{3/2}}.
\end{eqnarray*}

We only consider $\mathbb{P}(2\mid1)$. The analysis of $\mathbb
{P}(1\mid2)$
is similar. Suppose $\Xb\mid\break \mbox{class 1} \stackrel{(d)}{=}\Zb\sim
\mathcal{N}(\bmu_1,\bSigma_1)$. Define
\[
\Yb= \bSigma^{-1/2}_1(\Zb-\bmu_1),
\]
so that $\Yb\sim\mathcal{N}(\bzero,\bI_d)$ and $\Zb=\bSigma
^{1/2}_1\Yb+\bmu_1$.
Note that
%
%
\begin{eqnarray}
\label{temp1-errgap} Q(\Zb) &=& \bigl(\bSigma^{1/2}_1\Yb+
\bmu_1 \bigr)^{\top}\bOmega\bigl(\bSigma
^{1/2}_1 \Yb+\bmu_1 \bigr)-2 \bigl(
\bSigma^{1/2}_1\Yb+ \bmu_1
\bigr)^{\top}\bdelta
\nonumber\\[-8pt]\\[-8pt]\nonumber
&=&\Yb^{\top} \bSigma^{1/2}_1
\bOmega\bSigma^{1/2}_1 \bY+2\Yb^{\top
}
\bSigma^{1/2}_1(\bOmega\bmu_1-\bdelta)+
\bmu^{\top}_1\bOmega\bmu_1-2
\bmu^{\top}_1\bdelta.
\end{eqnarray}
Recall that $\bSigma^{1/2}_1\bOmega\bSigma^{1/2}_1=\bK_1\bS_1\bK
_1^{\top}$ is { the eigen-decomposition by excluding the $0$
eigenvalues. Since $\bSigma_1$ has full rank and the rank of $\bOmega
$ is $q$, the rank of $\bSigma^{1/2}_1\bOmega\bSigma^{1/2}_1$ is
$q$. Therefore,\vspace*{2pt} $\bS_1$ is a $q\times q$ diagonal matrix, and $\bK_1$
is a $d\times q$ matrix satisfying $\bK_1^{\top}\bK_1=\bI_q$.} Let
$\widetilde{\bK}_1$ be any {$d\times(d-q)$} matrix such that $\bK
=[\bK_1, \widetilde{\bK}_1]$ is a {$d\times d$} orthogonal matrix.
Since $\bI_d=\bK\bK^{\top}=\bK_1\bK_1^{\top}+\widetilde{\bK
}_1\widetilde{\bK}_1^{\top}$, we have
\[
\Yb^{\top}\bSigma^{1/2}_1(\bOmega
\bmu_1-\bdelta) =\Yb^{\top}\bK_1
\bK_1^{\top}\bSigma^{1/2}_1(\bOmega\bmu
_1-\bdelta) + \Yb^{\top}\widetilde{\bK}_1
\widetilde{\bK}_1^{\top
}\bSigma^{1/2}_1(
\bOmega\bmu_1-\bdelta).
\]
We\vspace*{1pt} recall that $\bbeta_1=\bK_1^{\top}\bSigma^{1/2}_1(\bOmega\bmu
_1-\bbeta)$.
Let $\widetilde{\bbeta}_1=\widetilde{\bK}_1^{\top}\bSigma
^{1/2}_1(\bOmega\bmu_1-\bdelta)$,
$\Wb=\bK_1^{\top}\Yb$, $\widetilde{\Wb}=\widetilde{\bK}_1^{\top
}\Yb$ and $c_1=\bmu^{\top}_1\bOmega\bmu_1-2\bmu^{\top}_1\bdelta
$. It follows from (\ref{temp1-errgap}) that
\begin{eqnarray*}
Q(\Zb) &=&\Yb^{\top}{\bK_1\bS_1
\bK_1^{\top}}\Yb+ 2\Yb^{\top
}\bK_1
\bbeta_1 + 2 \Yb^{\top}\widetilde{\bK}_1
\widetilde{\bbeta}_1 + c_1
\\
&=& \Wb^{\top}{
\bS_1}\Wb+2\Wb^{\top}\bbeta_1 + 2\widetilde{\Wb
}^{\top}\widetilde{\bbeta}_1 + c_1
\\
&\equiv&
\bar{Q}_1(\Wb)+ \bar{F}_1(\widetilde{\Wb}) +
c_1,
\end{eqnarray*}
where {$\bar{Q}_1(\bw)=\bw^{\top}\bS_1\bw+ 2\bw^{\top}\bbeta
_1$ and $\bar{F}_1(\bw)=2\bw^{\top}\widetilde{\bbeta}_1$}.
Therefore,
\[
\mathbb{P}(2\mid1)=\mathbb{P} \bigl(Q(\Zb)>c \bigr)=\mathbb{P} \bigl
(\bar
{Q}_1(\Wb) + \bar{F}_1(\widetilde{\Wb})>c-c_1
\bigr).
\]
We write for convenience $\Wb=(W_1,\ldots, W_q)^{\top}$, $\widetilde
{\Wb}=(W_{q+1},\ldots, W_d)^{\top}$, $\bbeta_1=(\beta_{11},\ldots
,\beta_{1q})^{\top}$ and $\widetilde{\bbeta}_1=(\beta_{1(q+1)},
\ldots, \beta_{1d})^{\top}$, and notice that $W_i\stackrel
{\mathrm{i.i.d.}}{\sim
}N(0,1)$ for $1\leq i\leq d$.
Moreover,
%
%
\begin{equation}
\label{errgap-temp} \bar{Q}_1(\Wb) + \bar{F}_1(\widetilde{
\Wb}) = \sum_{i=1}^{q} \bigl(s_i
W^2_i+2W_i\beta_{1i} \bigr)+
\sum_{i=q+1}^{d} 2W_i
\beta_{1i} \equiv\sum_{i=1}^{d}
\xi_i, 
\end{equation}
where $\xi_i=s_iW_i^2 I\{1\leq i\leq q\}+2W_i\beta_{1i}$, for $1\leq
i\leq d$.
The right-hand side of (\ref{errgap-temp}) is a sum of independent
variables, so we can apply the Edgeworth expansion to its distribution
function, as described in detail below.

Note that $\mathbb{E}(W_i^2)=1$, $\mathbb{E}(W_i^4)=3$, $\mathbb
{E}(W_i^6)=15$ and $\mathbb{E}(W_i^{2j+1})=0$ for nonnegative integers
$j$. By direct calculation,
\begin{eqnarray*}
\eta_1& \equiv& \sum_{i=1}^d
\mathbb{E}(\xi_i)=\sum_{i=1}^q
s_i=\tr(\bS_1)=\tr(\bOmega\bSigma_1),
\\
\eta_2& \equiv&\sum_{i=1}^d
\operatorname{var} (\xi_i) = \sum_{i=1}^q
\bigl( 2s_i^2 +4\beta^2_{1i}
\bigr) + \sum_{i=q+1}^d 4
\beta^2_{1i} = 2\tr\bigl(\bS_1^2
\bigr)+4 \llvert\bbeta_1 \rrvert^2 + 4 \llvert
\widetilde{\bbeta}_1\rrvert^2
\\
& =&2\tr(\bOmega
\bSigma_1\bOmega\bSigma_1)+4 (\bOmega\bmu
_1-\bdelta)^{\top}\bSigma_1(\bOmega
\bmu_1-\bdelta),
\\
\eta_3& \equiv&\sum
_{i=1}^d \mathbb{E} \bigl[\xi_i-
\mathbb{E}(\xi_i) \bigr]^3 = \sum
_{i=1}^d \bigl(8 s^3_i+24
\beta^2_{1i} s_i \bigr)
\\
&=&8\tr\bigl(
\bS_1^3 \bigr)+24 \bbeta_1^{\top}
\bS_1\bbeta_1 =8\tr\bigl[(\bOmega\bSigma_1)^3
\bigr]+24(\bOmega\bmu_1-\bdelta)^{\top
}\bSigma_1
\bOmega\bSigma_1(\bOmega\bmu_1-\bdelta).
\end{eqnarray*}
Notice that $\mathbb{E}(\llvert \xi_i-\mathbb{E}(\xi_i)\rrvert
^3)<\infty$, as
$\max\{\llvert s_i\rrvert,\llvert \beta_{1i}\rrvert, 1\leq i\leq d\}
\leq C_0$ by assumption.
Using results from Chapter XVI of \citet{Feller1966},
we know
\begin{eqnarray*}
\mathbb{P}(2\mid1)&=&\mathbb{P} \Biggl(\sum
_{i=1}^{d} \xi_i >c-c_1
\Biggr)
\\
&=&\mathbb{P} \biggl(\frac{\sum_{i=1}^d \xi_i-\mathbb{E}(\sum_{i=1}^d \xi
_i)}{\sqrt{\sum_{i=1}^d \operatorname{var}(\xi
_i)}}>\frac{c-c_1-\mathbb{E}(\sum_{i=1}^d \xi_i)}{\sqrt{\sum_{i=1}^d
\operatorname{var}(\xi_i)}} \biggr)
\\
&=&\bar{
\Phi} \biggl(\frac{c-c_1-\eta_1}{\sqrt{\eta_2}} \biggr) +\frac{\eta
_3(1-(\sfrac{(c_1-c+\eta_1)^2}{\eta_2}))}{6 \eta
^{3/2}_2}\phi\biggl(
\frac{c_1-c+\eta_1}{\sqrt{\eta_2}} \biggr)
\\
&&{} +o \biggl(\frac{d}{\eta
^{3/2}_2} \biggr),
\nonumber
\end{eqnarray*}
where $\phi$ is the probability density function of the standard
normal distribution. It is observed that $\eta_2=L_1(\bOmega,\bdelta
)$ and $c_1+\eta_1=M_1(\bOmega,\bdelta)$. Also, $c=tM_1(\bOmega
,\bdelta)+(1-t)M_2(\bOmega,\bdelta)$. As a result,
\begin{eqnarray*}
\frac{c-c_1-\eta_1}{\sqrt{\eta_2}} &=& \frac{[tM_1 + (1-t)
M_2]-M_1}{\sqrt{L_1}} = \frac{(1-t)(M_2 - M_1)}{\sqrt{L_1}}
\\
&=& (1-t)
\frac{M}{\sqrt{L_1}}.
\end{eqnarray*}
Plugging this into the expression of $\mathbb{P}(2\mid1)$, the first term
is $\bar{\Phi}((1-t)\tfrac{M}{\sqrt{L_1}})$. Moreover,
since the function $(1-u^2)\phi(u)$ is uniformly bounded, the second
term is $O(\frac{\eta_3}{\eta_2^{3/2}})$. Here $\eta_2=L_1$, and
$\eta_3=O(q)$ as $s_i$'s and $\beta_{1i}$'s are abounded in magnitude.
Combining the above gives
\[
\mathbb{P}(2\mid1) = \bar{\Phi} \biggl(\frac{(1-t)M}{\sqrt{L_1}} \biggr
) +
\frac{O(q) + o(d)}{L_1^{3/2}}.
\]
The proof is now complete.



\begin{supplement}[id=suppA]
\stitle{Supplement to ``QUADRO: A supervised dimension reduction method via Rayleigh quotient optimization''}
\slink[doi]{10.1214/14-AOS1307SUPP} 
\sdatatype{.pdf}
\sfilename{AOS1307\_supp.pdf}
\sdescription{Owing to space constraints, numerical tables for
simulation and some of the technical proofs are relegated to a
supplementary document. It contains proofs of Propositions \ref
{propellipmeanvar}, \ref{propQeigen} and \ref{propH-Taylor}.}
\end{supplement}

%

\printaddresses

\begin{thebibliography}{38}

\bibitem[\protect\citeauthoryear{Bickel, Ritov and Tsybakov}{2009}]{bickel2009simultaneous}
%
\begin{barticle}[mr]
\bauthor{\bsnm{Bickel},~\bfnm{Peter~J.}\binits{P.~J.}},
\bauthor{\bsnm{Ritov},~\bfnm{Ya'acov}\binits{Y.}} \AND
\bauthor{\bsnm{Tsybakov},~\bfnm{Alexandre~B.}\binits{A.~B.}}
(\byear{2009}).
\btitle{Simultaneous analysis of lasso and {D}antzig selector}.
\bjournal{Ann. Statist.}
\bvolume{37}
\bpages{1705--1732}.
\bid{doi={10.1214/08-AOS620}, issn={0090-5364}, mr={2533469}}
\bptnote{check volume}%
\end{barticle}
%

\bptok{imsref}%
\endbibitem

\bibitem[\protect\citeauthoryear{Bindea et~al.}{2009}]{bindea2009cluego}
%
\begin{barticle}[author]
\bauthor{\bsnm{Bindea},~\bfnm{Gabriela}\binits{G.}},
\bauthor{\bsnm{Mlecnik},~\bfnm{Bernhard}\binits{B.}},
\bauthor{\bsnm{Hackl},~\bfnm{Hubert}\binits{H.}},
\bauthor{\bsnm{Charoentong},~\bfnm{Pornpimol}\binits{P.}},
\bauthor{\bsnm{Tosolini},~\bfnm{Marie}\binits{M.}},
\bauthor{\bsnm{Kirilovsky},~\bfnm{Amos}\binits{A.}},
\bauthor{\bsnm{Fridman},~\bfnm{Wolf-Herman}\binits{W.-H.}},
\bauthor{\bsnm{Pag{\`e}s},~\bfnm{Franck}\binits{F.}},
\bauthor{\bsnm{Trajanoski},~\bfnm{Zlatko}\binits{Z.}} \AND
\bauthor{\bsnm{Galon},~\bfnm{J{\'e}r{\^o}me}\binits{J.}}
(\byear{2009}).
\btitle{ClueGO: A cytoscape plug-in to decipher functionally grouped
gene ontology and pathway annotation networks}.
\bjournal{Bioinformatics}
\bvolume{25}
\bpages{1091--1093}.
\end{barticle}
%

\bptok{imsref}%
\endbibitem

\bibitem[\protect\citeauthoryear{Cai and Liu}{2011}]{LPD}
%
\begin{barticle}[mr]
\bauthor{\bsnm{Cai},~\bfnm{Tony}\binits{T.}} \AND
\bauthor{\bsnm{Liu},~\bfnm{Weidong}\binits{W.}}
(\byear{2011}).
\btitle{A direct estimation approach to sparse linear discriminant analysis}.
\bjournal{J.~Amer. Statist. Assoc.}
\bvolume{106}
\bpages{1566--1577}.
\bid{doi={10.1198/jasa.2011.tm11199}, issn={0162-1459}, mr={2896857}}
\end{barticle}
%

\bptok{imsref}%
\endbibitem

\bibitem[\protect\citeauthoryear{Cai, Liu and Luo}{2011}]{Clime}
%
\begin{barticle}[mr]
\bauthor{\bsnm{Cai},~\bfnm{Tony}\binits{T.}},
\bauthor{\bsnm{Liu},~\bfnm{Weidong}\binits{W.}} \AND
\bauthor{\bsnm{Luo},~\bfnm{Xi}\binits{X.}}
(\byear{2011}).
\btitle{A constrained {$\ell\sb1$} minimization approach to sparse
precision matrix estimation}.
\bjournal{J. Amer. Statist. Assoc.}
\bvolume{106}
\bpages{594--607}.
\bid{doi={10.1198/jasa.2011.tm10155}, issn={0162-1459}, mr={2847973}}
\bptnote{check pages}%
\end{barticle}
%

\bptok{imsref}%
\endbibitem

\bibitem[\protect\citeauthoryear{Catoni}{2012}]{Catoni11}
%
\begin{barticle}[mr]
\bauthor{\bsnm{Catoni},~\bfnm{Olivier}\binits{O.}}
(\byear{2012}).
\btitle{Challenging the empirical mean and empirical variance: A
deviation study}.
\bjournal{Ann. Inst. Henri Poincar\'e Probab. Stat.}
\bvolume{48}
\bpages{1148--1185}.
\bid{doi={10.1214/11-AIHP454}, issn={0246-0203}, mr={3052407}}
\end{barticle}
%

\bptok{imsref}%
\endbibitem

\bibitem[\protect\citeauthoryear{Chen, Zou and Cook}{2010}]{chen2010coordinate}
%
\begin{barticle}[mr]
\bauthor{\bsnm{Chen},~\bfnm{Xin}\binits{X.}},
\bauthor{\bsnm{Zou},~\bfnm{Changliang}\binits{C.}} \AND
\bauthor{\bsnm{Cook},~\bfnm{R.~Dennis}\binits{R.~D.}}
(\byear{2010}).
\btitle{Coordinate-independent sparse sufficient dimension reduction
and variable selection}.
\bjournal{Ann. Statist.}
\bvolume{38}
\bpages{3696--3723}.
\bid{doi={10.1214/10-AOS826}, issn={0090-5364}, mr={2766865}}
\bptnote{check volume}%
\end{barticle}
%

\bptok{imsref}%
\endbibitem

\bibitem[\protect\citeauthoryear{Cook and Weisberg}{1991}]{cook1991comment}
%
\begin{barticle}[auto]
\bauthor{\bsnm{Cook},~\bfnm{R~Dennis}\binits{R.~D.}} \AND
\bauthor{\bsnm{Weisberg},~\bfnm{Sanford}\binits{S.}}
(\byear{1991}).
\btitle{Comment on ``Sliced inverse regression for dimension reduction.''}
\bjournal{J. Amer. Statist. Assoc.}
\bvolume{86}
\bpages{328--332}.
\end{barticle}
%

\bptok{imsref}%
\endbibitem

\bibitem[\protect\citeauthoryear{Coudret, Liquet and
Saracco}{2014}]{coudret2014comparison}
%
\begin{barticle}[mr]
\bauthor{\bsnm{Coudret},~\bfnm{Rapha{\"e}l}\binits{R.}},
\bauthor{\bsnm{Liquet},~\bfnm{Benoit}\binits{B.}} \AND
\bauthor{\bsnm{Saracco},~\bfnm{J{\'e}r{\^o}me}\binits{J.}}
(\byear{2014}).
\btitle{Comparison of sliced inverse regression approaches for
underdetermined cases}.
\bjournal{J. SFdS}
\bvolume{155}
\bpages{72--96}.
\bid{issn={2102-6238}, mr={3211755}}
\end{barticle}
%

\bptok{imsref}%
\endbibitem

\bibitem[\protect\citeauthoryear{Fan and Fan}{2008}]{FAIR}
%
\begin{barticle}[mr]
\bauthor{\bsnm{Fan},~\bfnm{Jianqing}\binits{J.}} \AND
\bauthor{\bsnm{Fan},~\bfnm{Yingying}\binits{Y.}}
(\byear{2008}).
\btitle{High-dimensional classification using features annealed
independence rules}.
\bjournal{Ann. Statist.}
\bvolume{36}
\bpages{2605--2637}.
\bid{doi={10.1214/07-AOS504}, issn={0090-5364}, mr={2485009}}
\end{barticle}
%

\bptok{imsref}%
\endbibitem

\bibitem[\protect\citeauthoryear{Fan, Feng and Tong}{2012}]{ROAD}
%
\begin{barticle}[mr]
\bauthor{\bsnm{Fan},~\bfnm{J.}\binits{J.}},
\bauthor{\bsnm{Feng},~\bfnm{Y.}\binits{Y.}} \AND
\bauthor{\bsnm{Tong},~\bfnm{X.}\binits{X.}}
(\byear{2012}).
\btitle{A road to classification in high dimensional space}.
\bjournal{J.~Roy. Statist. Soc. B}
\bvolume{74}
\bpages{745--771}.
\bid{mr={2965958}}
\end{barticle}
%

\bptok{imsref}%
\endbibitem

\bibitem[\protect\citeauthoryear{Fan and Li}{2001}]{fan2001variable}
%
\begin{barticle}[mr]
\bauthor{\bsnm{Fan},~\bfnm{Jianqing}\binits{J.}} \AND
\bauthor{\bsnm{Li},~\bfnm{Runze}\binits{R.}}
(\byear{2001}).
\btitle{Variable selection via nonconcave penalized likelihood and its
oracle properties}.
\bjournal{J. Amer. Statist. Assoc.}
\bvolume{96}
\bpages{1348--1360}.
\bid{doi={10.1198/016214501753382273}, issn={0162-1459}, mr={1946581}}
\end{barticle}
%

\bptok{imsref}%
\endbibitem

\bibitem[\protect\citeauthoryear{Fan, Xue and Zou}{2014}]{fan2014strong}
%
\begin{barticle}[mr]
\bauthor{\bsnm{Fan},~\bfnm{Jianqing}\binits{J.}},
\bauthor{\bsnm{Xue},~\bfnm{Lingzhou}\binits{L.}} \AND
\bauthor{\bsnm{Zou},~\bfnm{Hui}\binits{H.}}
(\byear{2014}).
\btitle{Strong oracle optimality of folded concave penalized estimation}.
\bjournal{Ann. Statist.}
\bvolume{42}
\bpages{819--849}.
\bid{doi={10.1214/13-AOS1198}, issn={0090-5364}, mr={3210988}}
\end{barticle}
%

\bptok{imsref}%
\endbibitem

\bibitem[\protect\citeauthoryear{Fan et~al.}{2014}]{QUADROsupp}
%
\begin{bmisc}[author]
\bauthor{\bsnm{Fan},~\bfnm{Jianqing}\binits{J.}},
\bauthor{\bsnm{Ke},~\bfnm{Zheng~Tracy}\binits{Z.~T.}},
\bauthor{\bsnm{Liu},~\bfnm{Han}\binits{H.}} \AND
\bauthor{\bsnm{Xia},~\bfnm{Lucy}\binits{L.}}
(\byear{2015}).
\bhowpublished{Supplement to ``QUADRO: A supervised dimension reduction
method via Rayleigh
quotient optimization.''
DOI:\doiurl{10.1214/14-AOS1307SUPP}}.
\bptok{imsref}%
\end{bmisc}
%
\endbibitem
%

\bptok{imsref}%
\endbibitem

\bibitem[\protect\citeauthoryear{Feller}{1966}]{Feller1966}
%
\begin{bbook}[author]
\bauthor{\bsnm{Feller},~\bfnm{William}\binits{W.}}
(\byear{1966}).
\btitle{An Introduction to Probability Theory and Its Applications.
Vol. II}.
\bpublisher{Wiley},
\blocation{New York}.
\end{bbook}
%

\bptok{imsref}%
\endbibitem

\bibitem[\protect\citeauthoryear{Fisher}{1936}]{fisher1936use}
%
\begin{barticle}[author]
\bauthor{\bsnm{Fisher},~\bfnm{Ronald~A.}\binits{R.~A.}}
(\byear{1936}).
\btitle{The use of multiple measurements in taxonomic problems}.
\bjournal{Annals of Eugenics}
\bvolume{7}
\bpages{179--188}.
\end{barticle}
%

\bptok{imsref}%
\endbibitem

\bibitem[\protect\citeauthoryear{Friedman}{1989}]{friedman1989regularized}
%
\begin{barticle}[mr]
\bauthor{\bsnm{Friedman},~\bfnm{Jerome~H.}\binits{J.~H.}}
(\byear{1989}).
\btitle{Regularized discriminant analysis}.
\bjournal{J. Amer. Statist. Assoc.}
\bvolume{84}
\bpages{165--175}.
\bid{issn={0162-1459}, mr={0999675}}
\end{barticle}
%

\bptok{imsref}%
\endbibitem

\bibitem[\protect\citeauthoryear{Guo, Hastie and Tibshirani}{2005}]{SCRDA}
%
\begin{barticle}[author]
\bauthor{\bsnm{Guo},~\bfnm{Y.}\binits{Y.}},
\bauthor{\bsnm{Hastie},~\bfnm{T.}\binits{T.}} \AND
\bauthor{\bsnm{Tibshirani},~\bfnm{R.}\binits{R.}}
(\byear{2005}).
\btitle{Regularized discriminant analysis and its application in microarrays}.
\bjournal{Biostatistics}
\bvolume{1}
\bpages{1--18}.
\end{barticle}
%

\bptok{imsref}%
\endbibitem

\bibitem[\protect\citeauthoryear{Han and Liu}{2012}]{HanLiu12}
%
\begin{barticle}[author]
\bauthor{\bsnm{Han},~\bfnm{F.}\binits{F.}} \AND
\bauthor{\bsnm{Liu},~\bfnm{H.}\binits{H.}}
(\byear{2012}).
\btitle{Transelliptical component analysis}.
\bjournal{Adv. Neural Inf. Process. Syst.}
\bvolume{25}
\bpages{368--376}.
\end{barticle}
%

\bptok{imsref}%
\endbibitem

\bibitem[\protect\citeauthoryear{Han, Zhao and Liu}{2013}]{Coda}
%
\begin{barticle}[mr]
\bauthor{\bsnm{Han},~\bfnm{Fang}\binits{F.}},
\bauthor{\bsnm{Zhao},~\bfnm{Tuo}\binits{T.}} \AND
\bauthor{\bsnm{Liu},~\bfnm{Han}\binits{H.}}
(\byear{2013}).
\btitle{C{ODA}: High dimensional copula discriminant analysis}.
\bjournal{J.~Mach. Learn. Res.}
\bvolume{14}
\bpages{629--671}.
\bid{issn={1532-4435}, mr={3033343}}
\end{barticle}
%

\bptok{imsref}%
\endbibitem

\bibitem[\protect\citeauthoryear{Jiang and Liu}{2013}]{jiang2013sliced}
%
\begin{bmisc}[author]
\bauthor{\bsnm{Jiang},~\bfnm{Bo}\binits{B.}} \AND
\bauthor{\bsnm{Liu},~\bfnm{Jun~S.}\binits{J.~S.}}
(\byear{2013}).
\bhowpublished{Sliced inverse regression with variable selection and
interaction detection.
Preprint. Available at \arxivurl{arXiv:1304.4056}.}
\end{bmisc}
%

\bptok{imsref}%
\endbibitem

\bibitem[\protect\citeauthoryear{Kendall}{1938}]{Kendall}
%
\begin{barticle}[author]
\bauthor{\bsnm{Kendall},~\bfnm{M.~G.}\binits{M.~G.}}
(\byear{1938}).
\btitle{A new measure of rank correlation}.
\bjournal{Biometrika}
\bvolume{30}
\bpages{81--93}.
\end{barticle}
%

\bptok{imsref}%
\endbibitem

\bibitem[\protect\citeauthoryear{Kent}{1991}]{kent1991discussion}
%
\begin{barticle}[author]
\bauthor{\bsnm{Kent},~\bfnm{J.~T.}\binits{J.~T.}}
(\byear{1991}).
\btitle{Discussion of Li (1991)}.
\bjournal{J. Amer. Statist. Assoc.}
\bvolume{86}
\bpages{336--337}.
\end{barticle}
%

\bptok{imsref}%
\endbibitem

\bibitem[\protect\citeauthoryear{Li}{1991}]{li1991sliced}
%
\begin{barticle}[mr]
\bauthor{\bsnm{Li},~\bfnm{Ker-Chau}\binits{K.-C.}}
(\byear{1991}).
\btitle{Sliced inverse regression for dimension reduction}.
\bjournal{J. Amer. Statist. Assoc.}
\bvolume{86}
\bpages{316--342}.
\bid{issn={0162-1459}, mr={1137117}}
\bptnote{check related, check pages}%
\end{barticle}
%

\bptok{imsref}%
\endbibitem

\bibitem[\protect\citeauthoryear{Li}{2000}]{Li00highdimensional}
%
\begin{bmisc}[author]
\bauthor{\bsnm{Li},~\bfnm{Ker-Chau}\binits{K.-C.}}
(\byear{2000}).
\bhowpublished{High dimensional data analysis via the SIR/PHD approach.
Lecture notes, Dept. Statistics, UCLA, Los Angeles, CA. Available at \surl{http://www.stat.ucla.edu/\textasciitilde kcli/sir-PHD.pdf}.}
\end{bmisc}
%

\bptok{imsref}%
\endbibitem

\bibitem[\protect\citeauthoryear{Li and Wang}{2007}]{li2007directional}
%
\begin{barticle}[mr]
\bauthor{\bsnm{Li},~\bfnm{Bing}\binits{B.}} \AND
\bauthor{\bsnm{Wang},~\bfnm{Shaoli}\binits{S.}}
(\byear{2007}).
\btitle{On directional regression for dimension reduction}.
\bjournal{J. Amer. Statist. Assoc.}
\bvolume{102}
\bpages{997--1008}.
\bid{doi={10.1198/016214507000000536}, issn={0162-1459}, mr={2354409}}
\end{barticle}
%

\bptok{imsref}%
\endbibitem

\bibitem[\protect\citeauthoryear{Li and Yin}{2008}]{li2008sliced}
%
\begin{barticle}[mr]
\bauthor{\bsnm{Li},~\bfnm{Lexin}\binits{L.}} \AND
\bauthor{\bsnm{Yin},~\bfnm{Xiangrong}\binits{X.}}
(\byear{2008}).
\btitle{Sliced inverse regression with regularizations}.
\bjournal{Biometrics}
\bvolume{64}
\bpages{124--131}.
\bid{doi={10.1111/j.1541-0420.2007.00836.x}, issn={0006-341X}, mr={2422826}}
\end{barticle}
%

\bptok{imsref}%
\endbibitem

\bibitem[\protect\citeauthoryear{Liu et~al.}{2012}]{liu2012high}
%
\begin{barticle}[mr]
\bauthor{\bsnm{Liu},~\bfnm{Han}\binits{H.}},
\bauthor{\bsnm{Han},~\bfnm{Fang}\binits{F.}},
\bauthor{\bsnm{Yuan},~\bfnm{Ming}\binits{M.}},
\bauthor{\bsnm{Lafferty},~\bfnm{John}\binits{J.}} \AND
\bauthor{\bsnm{Wasserman},~\bfnm{Larry}\binits{L.}}
(\byear{2012}).
\btitle{High-dimensional semiparametric {G}aussian copula graphical models}.
\bjournal{Ann. Statist.}
\bvolume{40}
\bpages{2293--2326}.
\bid{doi={10.1214/12-AOS1037}, issn={0090-5364}, mr={3059084}}
\end{barticle}
%

\bptok{imsref}%
\endbibitem

\bibitem[\protect\citeauthoryear{Luparello}{2013}]{luparello2013aspects}
%
\begin{barticle}[author]
\bauthor{\bsnm{Luparello},~\bfnm{Claudio}\binits{C.}}
(\byear{2013}).
\btitle{Aspects of collagen changes in breast cancer}.
\bjournal{J. Carcinogene Mutagene S13:007}.
\bnote{DOI:\doiurl{10.4172/2157-2518.S13-007}}.
\end{barticle}
%

\bptok{imsref}%
\endbibitem

\bibitem[\protect\citeauthoryear{Maruyama and Seo}{2003}]{kurtosis}
%
\begin{barticle}[mr]
\bauthor{\bsnm{Maruyama},~\bfnm{Yosihito}\binits{Y.}} \AND
\bauthor{\bsnm{Seo},~\bfnm{Takashi}\binits{T.}}
(\byear{2003}).
\btitle{Estimation of moment parameter in elliptical distributions}.
\bjournal{J.~Japan Statist. Soc.}
\bvolume{33}
\bpages{215--229}.
\bid{doi={10.14490/jjss.33.215}, issn={0389-5602}, mr={2039896}}
\end{barticle}
%

\bptok{imsref}%
\endbibitem

\bibitem[\protect\citeauthoryear{Shao et~al.}{2011}]{Shao11}
%
\begin{barticle}[mr]
\bauthor{\bsnm{Shao},~\bfnm{Jun}\binits{J.}},
\bauthor{\bsnm{Wang},~\bfnm{Yazhen}\binits{Y.}},
\bauthor{\bsnm{Deng},~\bfnm{Xinwei}\binits{X.}} \AND
\bauthor{\bsnm{Wang},~\bfnm{Sijian}\binits{S.}}
(\byear{2011}).
\btitle{Sparse linear discriminant analysis by thresholding for high
dimensional data}.
\bjournal{Ann. Statist.}
\bvolume{39}
\bpages{1241--1265}.
\bid{doi={10.1214/10-AOS870}, issn={0090-5364}, mr={2816353}}
\end{barticle}
%

\bptok{imsref}%
\endbibitem

\bibitem[\protect\citeauthoryear{Wei and Li}{2007}]{wei2007markov}
%
\begin{barticle}[pbm]
\bauthor{\bsnm{Wei},~\bfnm{Zhi}\binits{Z.}} \AND
\bauthor{\bsnm{Li},~\bfnm{Hongzhe}\binits{H.}}
(\byear{2007}).
\btitle{A Markov random field model for network-based analysis of
genomic data}.
\bjournal{Bioinformatics}
\bvolume{23}
\bpages{1537--1544}.
\bid{doi={10.1093/bioinformatics/btm129}, issn={1367-4811},
pii={btm129}, pmid={17483504}}
\end{barticle}
%

\bptok{imsref}%
\endbibitem

\bibitem[\protect\citeauthoryear{Witten and
Tibshirani}{2011}]{witten2011penalized}
%
\begin{barticle}[mr]
\bauthor{\bsnm{Witten},~\bfnm{Daniela~M.}\binits{D.~M.}} \AND
\bauthor{\bsnm{Tibshirani},~\bfnm{Robert}\binits{R.}}
(\byear{2011}).
\btitle{Penalized classification using {F}isher's linear discriminant}.
\bjournal{J. R. Stat. Soc. Ser. B. Stat. Methodol.}
\bvolume{73}
\bpages{753--772}.
\bid{doi={10.1111/j.1467-9868.2011.00783.x}, issn={1369-7412}, mr={2867457}}
\end{barticle}
%

\bptok{imsref}%
\endbibitem

\bibitem[\protect\citeauthoryear{Wu}{2008}]{wu2008kernel}
%
\begin{barticle}[mr]
\bauthor{\bsnm{Wu},~\bfnm{Han-Ming}\binits{H.-M.}}
(\byear{2008}).
\btitle{Kernel sliced inverse regression with applications to classification}.
\bjournal{J. Comput. Graph. Statist.}
\bvolume{17}
\bpages{590--610}.
\bid{doi={10.1198/106186008X345161}, issn={1061-8600}, mr={2528238}}
\bptnote{check pages}%
\end{barticle}
%

\bptok{imsref}%
\endbibitem

\bibitem[\protect\citeauthoryear{Zhao, Roeder and Liu}{2013}]{zhao2013psd}
%
\begin{bmisc}[author]
\bauthor{\bsnm{Zhao},~\bfnm{Tuo}\binits{T.}},
\bauthor{\bsnm{Roeder},~\bfnm{Kathryn}\binits{K.}} \AND
\bauthor{\bsnm{Liu},~\bfnm{Han}\binits{H.}}
(\byear{2013}).
\bhowpublished{Positive semidefinite rank-based correlation matrix
estimation with application to semiparametric graph estimation.
Unpublished manuscript.}
\end{bmisc}
%

\bptok{imsref}%
\endbibitem

\bibitem[\protect\citeauthoryear{Zhao and Yu}{2006}]{zhao2006model}
%
\begin{barticle}[mr]
\bauthor{\bsnm{Zhao},~\bfnm{Peng}\binits{P.}} \AND
\bauthor{\bsnm{Yu},~\bfnm{Bin}\binits{B.}}
(\byear{2006}).
\btitle{On model selection consistency of {L}asso}.
\bjournal{J. Mach. Learn. Res.}
\bvolume{7}
\bpages{2541--2563}.
\bid{issn={1532-4435}, mr={2274449}}
\end{barticle}
%

\bptok{imsref}%
\endbibitem

\bibitem[\protect\citeauthoryear{Zhong et~al.}{2005}]{zhong2005rsir}
%
\begin{barticle}[pbm]
\bauthor{\bsnm{Zhong},~\bfnm{Wenxuan}\binits{W.}},
\bauthor{\bsnm{Zeng},~\bfnm{Peng}\binits{P.}},
\bauthor{\bsnm{Ma},~\bfnm{Ping}\binits{P.}},
\bauthor{\bsnm{Liu},~\bfnm{Jun~S.}\binits{J.~S.}} \AND
\bauthor{\bsnm{Zhu},~\bfnm{Yu}\binits{Y.}}
(\byear{2005}).
\btitle{RSIR: Regularized sliced inverse regression for motif discovery}.
\bjournal{Bioinformatics}
\bvolume{21}
\bpages{4169--4175}.
\bid{doi={10.1093/bioinformatics/bti680}, issn={1367-4803},
pii={bti680}, pmid={16166098}}
\end{barticle}
%

\bptok{imsref}%
\endbibitem

\bibitem[\protect\citeauthoryear{Zou}{2006}]{zou2006adaptive}
%
\begin{barticle}[mr]
\bauthor{\bsnm{Zou},~\bfnm{Hui}\binits{H.}}
(\byear{2006}).
\btitle{The adaptive lasso and its oracle properties}.
\bjournal{J. Amer. Statist. Assoc.}
\bvolume{101}
\bpages{1418--1429}.
\bid{doi={10.1198/016214506000000735}, issn={0162-1459}, mr={2279469}}
\end{barticle}
%

\bptok{imsref}%
\endbibitem

\bibitem[\protect\citeauthoryear{Zou and Li}{2008}]{zou2008one}
%
\begin{barticle}[mr]
\bauthor{\bsnm{Zou},~\bfnm{Hui}\binits{H.}} \AND
\bauthor{\bsnm{Li},~\bfnm{Runze}\binits{R.}}
(\byear{2008}).
\btitle{One-step sparse estimates in nonconcave penalized likelihood models}.
\bjournal{Ann. Statist.}
\bvolume{36}
\bpages{1509--1533}.
\bid{doi={10.1214/009053607000000802}, issn={0090-5364}, mr={2435443}}
\bptnote{check pages}%
\end{barticle}
%

\bptok{imsref}%
\endbibitem

\end{thebibliography}
\end{document}